\documentclass[12pt,preprint]{aastex}
\usepackage{natbib}
\begin{document}

\title{The Environment on few Mpc scales of Infrared Luminous Galaxies at Redshifts $z$$\sim$1.}
\shorttitle{The Environment of Infrared Galaxies at $z$$\sim$1.}
\shortauthors{Marcillac et al.}

\author{
D.~Marcillac \altaffilmark{1},
G.~H.~Rieke\altaffilmark{1},
C.~Papovich\altaffilmark{1,2},
C.~N.~A. ~Willmer\altaffilmark{1},
B.~J.~Weiner\altaffilmark{1},
A.~L.~Coil \altaffilmark{1,3},
M.~C.~Cooper\altaffilmark{1,2,4}
B.~F.~Gerke\altaffilmark{4},
J.~Woo \altaffilmark{5},
J.~A.~Newman\altaffilmark{3,6},
A.~Georgakakis\altaffilmark{7},
E.~S.~Laird\altaffilmark{7},
K.~Nandra\altaffilmark{7},
G.~G.~Fazio\altaffilmark{9},
J.-S.~Huang\altaffilmark{9},
D.~C.~Koo\altaffilmark{8}}

\altaffiltext{1}{Steward Observatory, University of Arizona, 933
  N. Cherry Avenue, Tucson, AZ~85721}
\altaffiltext{2}{Spitzer Fellow}
\altaffiltext{3}{Hubble Fellow}
\altaffiltext{4}{Department of Astronomy, University of California,
Berkeley, CA 94720 -- 3411} 
\altaffiltext{5}{Racah Institute of Physics, The Hebrew University, Jerusalem, Israel} 
\altaffiltext{6}{Institute for Nuclear and Particle Astrophysics,
  Lawrence Berkeley National Laboratory, Berkeley, CA 94720} 
\altaffiltext{7}{Astrophysics Group, Blackett Laboratory, Imperial College, Prince, Consort Rd , London SW7 2BZ, UK}
\altaffiltext{8}{University of California Observatories/Lick
Observatory, Department of Astronomy and Astrophysics, University of
California, Santa Cruz, CA 95064}
\altaffiltext{9}{Harvard-Smithsonian Center for Astrophysics, 60
Garden Street, Cambridge, MA 02138}


\begin{abstract}

We investigate the environment of infrared luminous galaxies (L$_{IR}$[8-1000 $\mu $m$] >10^{11}$L$_{\sun}$). 
We focus on the redshift range 0.7 $\leq$ z $\leq$ 1, where 
these galaxies dominate the star formation activity and play a significant role in galaxy evolution. 
We employ MIPS 24$\mu$m data to identify infrared galaxies in the Extended Groth Strip (EGS). We use a local 
density indicator to probe the environment on few Mpc scales and a group member catalog, both of which make use of the DEEP2 spectroscopic redshift 
catalog, to quantify the environment of these galaxies.

We find that the local environment of LIRGs and ULIRGs is intermediate between that of blue and red galaxies.  
LIRGs and ULIRGs avoid underdense environments and inhabit local environments that are more dense on average 
than those of other DEEP2 galaxies at similar redshifts. However, when the comparison sample of the non-IR DEEP2 galaxies 
is restricted to have the same range of stellar mass, color, or luminosity as the IR--galaxies, there
is no longer any significant difference in environment; the IR-galaxies follow the same trends in 
the color-environment and luminosity-environment relations observed at z$\sim$1. 

We also find that about 30\% of the LIRGs and ULIRGs belong to groups, associated with a minimum dark 
matter halo of 6$\times$10$^{12}$M$_{\odot}$h$^{-1}$. The group members constitute 20 \% of the sources 
responsible for the IR star formation rate density and comoving energy density at z$\sim$1.

\end{abstract}

\section{Introduction}

The link between galaxy properties and environment has been studied extensively in the local Universe. 
For example, Oemler (1974) found that the late type galaxy fraction decreases with radius within a cluster, 
which is now known as the ``{\it radius-morphology}'' relation. Dressler (1980) extended this result to the 
``{\it density-morphology}'' relation, showing that the fraction of morphological types 
is a function of the density in the local environment.
This relation is found to hold in group and under-dense regions (Davis \& Geller, 1976, Postman \& Geller, 
1984) and it appears to operate over several orders of magnitude in local density.
Environment and galaxy properties such as colors, luminosity and physical parameters have also 
been studied thoroughly thanks to local surveys such as the Sloan Digital Sky Survey (SDSS, 
Blanton et al., 2003; Tanaka et al, 2004; Park et al., 2007). Gomez et al. (2003) show that the star 
formation rate (SFR) of galaxies is strongly correlated with the 
local projected galaxy density and defined a {\it{density-SFR}} relation, analogous to the 
{\it{density - morphology}} one. Clustering studies have given similar results (Zehavi et al. 
2002 for the SDSS, and for the 2dF Galaxy Redshift Survey (2dFGRS), Colless et al., 2001; Norberg 
et al., 2001). Balogh et al. (2004) also show that the relative numbers of star-forming and 
quiescent galaxies vary strongly as a function of local density.

The link between environment and galaxy properties is now being extended to the distant Universe, thanks to deep 
multi-wavelength redshift surveys such as VVDS (VIMOS VLT Deep Survey, Lefevre et al., 2004), DEEP2 (Deep 
Extragalactic Evolutionary Probe, Davis et al., 2003, 2005, Faber et al., 
in prep.) and COSMOS (Scoville et al., 2006). The majority of the relations or trends found in the local 
universe have now been extended to higher z. Ilbert et al. (2006) found there is a strong dependence of the 
luminosity function on large scale environment up to z$\sim$1.2. Cooper et al. (2006) show that the relations 
found in the local universe are still valid at z$\sim$1 : the local environment for red galaxies is more 
than 1.5 times denser than for blue ones. Galaxies with small [OII] equivalent width also occupy regions 
of higher density. The relationship between galaxy over-density and rest-frame color at z$\sim$1 also 
mirrors that seen in the local Universe. Cooper et al. (2007) find that the relationship between specific 
star formation rate and environment found at z$\sim$1 is similar to the one found locally. Coil et al. 
(2007) studied the color and luminosity dependence of galaxy clustering at z$\sim$1 and found similar results.

However, Elbaz et al. (2007) and Cooper et al. (2007) show that the relation between the total SFR and 
environment at z$\sim$1 is inverted from that observed at z$\sim$0.1 : while in the local universe, 
the mean SFR of galaxies decreases in regions of higher over-density, at z$\sim$1 it increases with 
local galaxy density. The nature of the infrared galaxy 
population also changes with redshift. 
In the local Universe, galaxies radiate $\sim$30\% of their bolometric luminosity over the 8-1000 micron 
range (Soifer \& Neugebauer, 1991), but LIRGs\footnote{LIRGs are classified as sources with $10^{11}\leq 
L_{IR} < 10^{12}$L$_{\sun}$.} and ULIRGs\footnote{ ULIRGs are classified as sources with 10$^{12}$L$_{\sun}$ 
$\leq$  L$_{IR}$.} are nonetheless rare locally and radiate 
only 6\% of the integrated IR emission of local galaxies and $\sim$ 2\% of the local galaxy total bolometric 
luminosity (Soifer \& Neugebauer, 1991, 
Sanders \& Mirabel, 1996). However, this population has strongly evolved in number or luminosity at z$\sim$1, 
where LIRGs are responsible for 70\% of the comoving IR energy density; they dominate the star forming 
activity beyond z$\sim$0.7 (Le Floc'h et al. 2005, P\'erez-Gonz\'alez et al. 2005). An important question 
is the link between LIRG/ULIRG evolution and the SFR-environment relation observed at z$\sim$1.
 
In addition, we do not understand what causes the strong evolution of LIRGs and ULIRGs. 
Gas exhaustion undoubtedly plays a role (Bell et al., 2005). Galaxy environment may also be 
important. The connection of IR activity to galaxy morphology or merging 
has often been proposed 
for both local and distant LIRGs and ULIRGs (Hammer et al., 2005, Sanders et al., 1988). The local
LIRGs and ULIRGs usually appear to have disturbed 
morphologies (Sanders \& Mirabel, 1996). Ishida (2004) study a sample of 56 local LIRGs drawn from the 
IRAS Bright Galaxy Sample (Soifer et al., 1987) and find that all galaxies with log (L$_{IR}$/L$_{\odot}$) 
$>$ 11.5 show at least some evidence of tidal features. The majority of galaxies 
with 11.1 $<$log (L$_{IR}$/L$_{\odot}$) $<$ 11.5 also tend to be disturbed, but $\sim$ 30\% are 
bright isolated spirals. In rough agreement, Wang et al. (2006) find that the 
fraction of interacting-merging galaxies and normal disk galaxies are respectively 48\% and 40\% among a 
sample of 159 local LIRGs detected in both the IRAS survey and the SDSS. 

However, there are intriguing clues that the morphologies of IR galaxies may be different at {\it{z}} $\sim$ 1.
Several independent studies have shown that LIRGs tend to be associated with massive spirals at z$>$0.6 
rather than with interacting sources 
(Bell et al., 2005, Melbourne et al., 2005, Bridge et al., 2007, Lotz et al., 2006). However, 
Shi et al. (2006) used the very deep images from the UDF (HST Ultra Deep Field survey, Beckwith 
et al, 2005) and found that a significant fraction of the disk-dominated LIRGs show some signs of 
morphological disturbance or asymmetry. Bridge et al. (2007) also found that a significant fraction 
of sources classified as disk-dominated LIRGs show signs 
of disturbance or are in pairs. Taken together, these results suggest that the morphological 
properties of LIRGs may change with redshift.

Another approach is to test whether the LIRG properties arise from environmental effects, e.g.,
the density-SFR relation. Studies of the environment of LIRGs have not been very extensive, but
they raise troubling questions.  
If so many local 
LIRGs and ULIRGs are triggered by galaxy interaction, then we might expect to find them 
predominantly in denser environments, groups or structures. However, this expectation is apparently 
not confirmed by the observations. Gonzalez-Solares et al. (2004) find 
a correlation length for the 3D infrared galaxy-galaxy autocorrelation function of r$_0$$\sim$4.3h$^{-1}$Mpc 
in the ELAIS (The European Large Area ISO Survey, Oliver et al., 2000, Rowan-Robinson et al., 2004) 
where \={z}$\sim$0.2, consistent with the previous findings from IRAS 
(Saunders et al., 1992, Fisher et al., 1994). These results indicate weaker 
autocorrelations than are found for optical galaxies in the same redshift range from 2dFGRS or 
SDSS (Norberg et al., 2001, Zehavi et al., 2002). Owers et al. (2007) also find that starbursting 
sources tend to be less clustered on large scales (5-15 Mpc) compared to the overall 2dFGRS galaxy 
population. However, the local samples used in both cases are biased toward galaxies with  L$_{IR}$ 
below the LIRG range, and they could have different clustering properties (Gilli et al., 2007). 

The environment of LIRGs and ULIRGs has not been studied extensively 
in the distant universe. Some works suggest that interactions might enhance IR activity 
at z$\sim$1 : Cohen et al., (2000) focused on sources in the HDFN (Hubble Deep Field North, 
Williams et al., 1996) detected at 15$\mu$m with ISO (\={z}$\sim$0.63) and showed 
that they are more clustered than average, with 90\% of them lying within the redshift peaks. 
Lin et al. (2007) also find a 
mid-IR enhancement in pairs separated by less than a few tens of kpc in the EGS (Extended Groth Strip) region.
Bridge et al. (2007) state that galaxies involved in a ``merger process''\footnote{In close pairs, or 
classified as merging-interacting systems by visual inspection or using the CAS parameters 
(Conselice et al., 2003a, 2003b)} 
account for $\sim$60\% of the IR luminosity density at z$\sim$1. However, none of these studies relies 
on a precise and systematic estimate of the galaxy local environment but rather on morphological or pair classification.

In this paper, we take advantage of the MIPS data obtained in the EGS in combination with other datasets 
in AEGIS (All-wavelength Extended Groth Strip International Survey, Davis et al., 2007) to study the 
local environment of LIRGs and ULIRGs 
at 0.7 $<$ z$<$ 1. 
We will quantify the mean environment of these sources disregarding their morphology or pair classification. 
We explore whether their environment is significantly different compared to the other R $<$ 24.1 galaxies 
at similar redshifts. In \S 2 we present the multi-wavelength data and environment catalogs 
that will be used through the paper. 
\S 3 presents the sample and its completeness limit and discusses how representative it is of the IR population.  
\S 4 addresses the issue of AGN contamination. 
\S 5 explores the SFR-density relation and expands the discussion of local environmental dependence for IR galaxies and other samples, 
and \S 6 address the link between IR galaxies and group membership.
\S 7 discusses our results, which are summarized in \S 8. 
Throughout this paper, we assume H$_o$= 70 km s$^{-1}$
 Mpc$^{-1}$, $\Omega_{\rm matter}$= 0.3 and $\Omega_{\Lambda}= 0.7$; all the magnitudes are in the AB system; all the galaxy masses are stellar masses unless otherwise is stated.

\section{The Data }
This paper utilizes a subset of the AEGIS data described in Davis et al. (2007). The data sets relevant to our work are described in this section.

\subsection{X-ray data}
 The EGS region has been observed by 8 ACIS-I pointings (Advanced CCD Imaging Spectrometer on {\it Chandra}) 
over $\sim$0.65deg$^2$ with an exposure time of 200ks per pointing.
The X-ray data reduction, source detection and flux estimation are carried out using methods 
described in Nandra et al. (2005). The 5-$\sigma$ detection levels are 1.1 and 8.2 $\times
10^{-16} erg s^{-1} cm^{-2}$ respectively in the soft (0.5 - 2.0 keV) and hard (2.0 - 7 keV) bands.

\subsection{Spitzer/MIPS 24 $\mu$m data}
The EGS was observed in January and June 2004 with MIPS (Multi-band Infrared Photometer for Spitzer, Rieke et al., 2004) 
on board the Spitzer Space Telescope as part of the {\it Spitzer} Guaranteed Time Observation (GTO) program. 
The observations were in the Slow Scan mode, which allows the coverage 
of large sky areas with high efficiency. The effective integration time at 24 $\mu$m 
is $\sim$1500 s in the deepest part of the field, which is the main focus of this paper.
The data were reduced with the MIPS Data Analysis Tool (DAT, Gordon et al. 2005). The final 
mosaic has an interpolated pixel size of 1.245 $\arcsec$ (half the native pixel scale) 
and covers 2.5$\times$0.2$\, \deg^2$.
The point-spread function (PSF) at 24 $\mu$m is $\sim$6 $\arcsec$ in diameter and hence 
virtually all the detected sources are unresolved. 
The 24 $\mu$m catalog utilized PSF-fitting photometry done with DAOPHOT (Stetson, 1987), 
with procedures described in Papovich et al. (2004).

We use data from the deepest part of the field (0.35 deg$^2$), which coincides with the DEEP2 redshift survey.
We used the approach of Papovich et al. (2004) to estimate the completeness of the data. 
Artificial sources were added to the images and extracted in the same way as real sources. 
The 50 and 80\% completeness limits are estimated to be 50 and 82 $\mu$Jy respectively. 
Papovich et al. (2004) also estimate the number of spurious sources as a function of 24 $\mu$m flux 
density in the Chandra Deep Field South, where their data have a similar total effective 
integration time and background as ours in the EGS. 
They show that at 82 $\mu$Jy, the false detection rate should be less than 3\% (see their Figure 1).
The final astrometry has been checked against the DEEP2 photometric catalog of Coil et al. (2004). A 
systematic offset of 0.1$\arcsec$ was found and removed. The remaining scatter 
relative to the DEEP2 photometric positions is estimated to be 0.6$\arcsec$ at 1$\sigma$. 
Our final catalog includes 4417 sources with f$_{24}>$ 82 $\mu$Jy.

\subsection{Spitzer/IRAC data}

IRAC (Fazio et al. 2004) data were also obtained under the GTO program in 2003 December
and 2004 June/July. Dithered exposures in all four IRAC bands (3.6, 4.5, 5.8 and 8 $\mu$m) 
were taken at 52 positions for net exposures of $\sim$ 2.7 hours per pointing. The data were
initially processed to the basic calibrated stage using version 11 of the Spitzer Science Center
pipeline. Custom IDL scripts were used to distortion-correct the individual frames, reject
cosmic rays, remove other image artifacts, and project the frames onto a common reference frame
for mosaicing. Identification of artifacts was facilitated because there were two data
sets obtained at position angles rotated by $\sim$ 180$^{\rm o}$. The final mosaic is sub-sampled
at 0.6${''}$ per pixel, half the native pixel size. 

DAOPHOT/FIND was used to identify sources in the mosaics and to obtain their photometry. 
Because of source crowding, neighboring objects in a surrounding 200-pixel box were removed
prior to obtaining photometry of a source in a 3$^{''}$ aperture. 
After applying aperture corrections, the source measurements
were entered into two catalogs, one selected at 3.6$\mu$m and containing about 73,000 objects,
and the other selected at 8$\mu$m with about 16,000 objects. The 5-$\sigma$ detection limits
are about 0.9, 0.9, 6.3, and 5.8$\mu$Jy respectively at 3.6, 4.5, 5.8 and 8 $\mu$m.

\subsection{Spectroscopic Redshift Catalog}

\label{deimos}
The EGS has been observed spectroscopically as part of the DEEP2  
collaboration (Davis et al., 2003, 2007). 
Approximately 13500 spectra have been obtained so far in the EGS 
with DEIMOS, the multi-object spectrograph on the Keck II telescope 
(Faber et al.,  2003).
About 9500 of these redshifts are secure ($>$95\% confidence),
defined as having at least two spectral features detected at a  
significant level.

Details of the DEEP2 observations, catalog construction and
data reduction can be found in Davis et al. (2003), Coil et al. (2004) and Davis et al. (2005). 
Sources targeted for spectroscopy are required to have 18.5 $\leq$ R$\leq$  
24.1 and surface brightness $\mu_R \le 26.5$. Stars and galaxies are distinguished 
using a combination of angular size, magnitude and
position in color-color space; objects chosen for spectroscopy have at least
a 20\% probability of being a galaxy based on their
photometry.
Not all sources matching these criteria can be observed on 
slitmasks, due to their high number density. Therefore, various
selection rules plus a final random selection are applied to 
pick sources for spectroscopy. The selection rules include 
a weight that is a function of both the R  
magnitude and the probability of being
a galaxy.  
This color cut is used in the other (non-AEGIS) DEEP2 fields to select  
galaxies at
z$>$0.75 for spectroscopy.  In the AEGIS field, galaxies at z$<$0.75 are  
also selected but
with a lower probability than a strict flux-limit would allow, which  
ensures a more uniform
redshift distribution that is not peaked at z=0.5. As a result, the  
observed density of
galaxies at z$>$0.75 in the AEGIS field is similar to other DEEP2 fields.

The DEEP2 survey is R-band-selected, which probes the rest-frame 
4000 \AA$\,$ at z$\sim$0.7 and $\sim$ 3400\AA$\,$ at z$\sim$1.
Therefore, at high redshift blue galaxies are preferentially selected. A 
bias against MIPS sources could result, since they are  
very dusty. However, part of this color bias is  
compensated by the intrinsic brightness of LIRGs and ULIRGs, which are  
among the brightest sources in this redshift range as a result of their 
strong star formation activity.
The selection effects and biases in the DEEP2
data are discussed extensively in Gerke et al. (2005). 
Here we focus only on the 0.7$ \leq $z$ \leq $1 redshift range, where these effects  
have been shown to be small.

\subsection{Photometric Redshift Catalog}
The EGS is also covered by one of the four fields from the Canada-France-Hawaii Telescope 
Legacy Survey (CFHTLS). The reduction and photometry of these images 
are described in McCracken et al. (2007, in preparation). The resulting catalog reaches a limiting magnitude 
of i$'_{AB}$$\sim$26. Photometric redshifts\footnote{available on the Cencos database (Le Brun et al. 
2007) at ${ \rm http://cencos.oamp.fr/cencos/CFHTLS/}$.} 
have been derived for objects with i$'\le$24 (Ilbert et al., 2006) using the {\it{Le Phare}} 
redshift code\footnote{$ \rm{www.lam.oamp.fr/arnouts/LE\_PHARE.html}$}, which provides a robust 
calibration method based on an iterative zero-point refinement combined with a template 
optimisation procedure and a Bayesian approach. 
VVDS provided thousands of redshifts to calibrate the photometric 
redshifts.
They have an accuracy of $\sigma$$_{\Delta z/(1+z)}$=0.029 with 
only 3.8\% of the sources with i'$_{AB}\le$ 24 having $ \rm{abs}(\Delta z$/(1+z))$\ge$0.15 
over the 0.2$ < $z$ < $1.5 redshift range. Although useful to test for completeness, these redshifts are not accurate enough to be used to study the local environment of galaxies; as a comparison, 
abs$({\Delta z/(1+z))_{DEEP2}}$=6.5$\times$10$^{-5}$ at z$\sim$1.

\subsection{Environment Characterization}

We characterize the environment of LIRGs and ULIRGs in two ways. The first is to estimate the 
local over-density around galaxies (1+$\delta_3$, Cooper et al. 2005), which we use to probe whether 
IR-luminous galaxies tend to be located in over-dense regions compared with field galaxies. The second is to examine whether LIRGs inhabit preferentially galaxy groups (Gerke et al. 2005, 2007). These approaches
give complementary insights about the environment of IR-luminous galaxies.

\subsubsection{Local over-density estimator}
\label{delta3}
We use the local over-density estimator described by Cooper et al. (2005, 2006). For each galaxy, 
a velocity interval of $\pm$1000km s$^{-1}$ is defined around the 
object along the line of sight; galaxies outside this interval are excluded as foreground
or background objects. All galaxies in the selected
redshift range are projected to the redshift of the source. The local galaxy surface density, $\Sigma$3, is 
defined as the projected third-nearest-neighbor surface density. However this estimation only 
includes sources with accurate spectroscopic redshifts in DEEP2; $\Sigma$3 varies as a function of 
redshift since both the sampling and the spectroscopic success and failures are functions of redshift. 
Therefore, for each density, Cooper et al. (2005) corrected $\Sigma$3 for the variable redshift 
completeness of the DEEP2 survey and for the redshift dependence of the sampling rate.
This correction is made by dividing each $\Sigma$3 by the median $<$$\Sigma$3$>$ of galaxies at 
this redshift in a redshift bin of $\delta$z=0.04. Cooper et al. (2005) have verified the validity of 
this empirical correction on the DEEP2 Mock galaxy catalogs of Yan, White and Coil (2004). Dividing $\Sigma$3 by the median value 
of $<$$\Sigma$3$>$ leads to an estimate of the over-density relative to the mean density at a given redshift 
and will be noted as 1+$\delta_3$ hereafter.

To minimize the impact of the survey edge, which could lead to biases in the local density estimator 
and properties, all galaxies within 1 h$^{-1}$ co-moving Mpc of an edge of the surveys 
have been removed from our samples.

\subsubsection{Group Catalog}
\label{group}

Groups are defined here as bound structures, 
virialized associations of two or more observed galaxies. 
The algorithm to identify groups is presented in Gerke et al. (2005). 
The groups are identified using the Voronoi-Delaunay Method (VDM) originally implemented 
in Marinoni et al. (2002). The group-finding algorithm proceeds iteratively. In the first
step, it identifies galaxies with a high density of neighbors; it next expands to a larger
volume to compute a richness for the possible group. Finally, the richness
estimate is corrected to account for the redshift-dependent number density of galaxies in the initial
survey. This algorithm has also been calibrated using the mock catalogs of Yan, White and Coil (2004), which are designed to replicate the DEEP2 survey. 
Working with the mock catalogs, the algorithm is subject to significant systematic errors for putative
groups with small velocity dispersions. However, it reproduces well the distribution 
of groups in redshift and velocity dispersion for velocity dispersions greater 
than 350 km s$^{-1}$ (Gerke et al. 2005). In this paper we 
used a recent group listing from Gerke et al. (2007), where the algorithm was applied to a large dataset.

We will compare the IR properties of galaxies belonging to two samples, one in 
groups and the other in the field. The first one, hereafter {\it{the group 
sample}}, contains the galaxies belonging to groups with velocity dispersions greater than 
100 km s$^{-1}$. We have decreased the velocity dispersion threshold from 350 km s$^{-1}$ to 100 km s$^{-1}$ 
and use the same prescription as Gerke et al. (2007) because of our emphasisis on the properties of the 
galaxies and not the group properties. This new velocity dispersion 
threshold has been chosen because below it, the number of false detections of putative groups 
increases significantly. Gerke et al. (2005) estimate that the fraction of real group members that are successfully 
identified is $\sim$79\%. This fraction does not vary much with velocity dispersion. The fraction of galaxies 
that are misidentified as group members in the 
resulting catalog is $\sim$46\%. This level of contamination seems to be high; however, tests with mock 
catalogs show that the majority of galaxies in our group catalog truly belong to groups. In addition, all the 
results presented in section~\ref{envvv} would only be stronger with a less contaminated sample.

The meaning of ``group'' needs to be clarified. ``Group members'' means at least two group galaxies 
are detected in the DEEP2 sample. If we take into account the DEEP2 spectroscopic completeness limit, 
a very approximate real richness could be about 4 members; this estimate does not take 
into account errors by the group-finder algorithm.
Coil et al. (2006) calculate the auto-correlation function of DEEP2 groups and estimate a minimum 
dark matter halo mass of 6$\times$10$^{12}$M$_{\odot}$h$^{-1}$ for them. 
Assuming that the dark matter halo growth is close to a factor 2 between z$\sim$1 and z$\sim$0 (Sheth and 
Tormen, 1999) and using the dark matter halo mass -- stellar mass relation that Yang et al. (2007) derived 
at z $\sim$0 using SDSS galaxy groups, we find that the stellar mass of the dark matter halos of the DEEP2 
groups is about 2$\times$10$^{11}$M$_{\odot}$h$^{-1}$ at z$\sim$0. This corresponds to a total stellar 
mass of $\sim$ five M$_{\star}$ galaxies, assuming M$_{\star}$=4$\times$10$^{10}$M$_{\odot}$h$^{-1}$ 
(Bell et al., 2003).   
 
The second sample, the {\it{field sample}} hereafter, is made of galaxies that do not belong to 
groups or belong to groups with a lower dispersion limit. Gerke et al. (2005) estimated that the fraction 
of misclassified group members represents only 6\% of the reconstructed field sample, showing 
that the field sample is in fact dominated by field galaxies.

\section{Defining a sample of LIRGs and ULIRGs.}
We define here the sample of LIRGs and ULIRGs used in this paper and address how complete and representative it is.

\subsection{Optical identification}
The full MIPS 24 $\mu$m catalog has 4417 sources detected with f$_{24} >$ 82 $\mu$Jy (80\% completeness limit) 
in the area where DEEP2 
spectroscopy is available. We cross-correlate the MIPS catalog with the DEEP2 photometric 
catalog (Coil et al., 2004) for galaxies with magnitudes R$\leq$24.1, using a 
search radius of 2$\arcsec$. This step ensures a unique association 
between a 24 $\mu$m source and a secure spectroscopic redshift and avoids misassociations arising from
the smaller number of sources in the redshift catalog.
The search radius accounts for local astrometric uncertainties between the optical and 24$\mu$m catalogs 
and allows for shifts in the brightest position of optical emission relative to the MIR peak 
(e.g., as in the Antennae galaxy, Mirabel et al., 1998). In the association process between 
the DEEP2 catalog and the MIPS sources, we find 2527 24 $\mu$m sources  with unique counterparts 
down to R=24.1. We rejected 219 associations 
because there was no unique optical identification. Of the 2527 associations, 1671 sources have a 
spectroscopic redshift with z$_q\geq$3 corresponding to 38\% of the total 
sample of MIPS sources and 66\% of MIPS sources with an optical counterpart. See Table~\ref{mips} 
for a summary of the different steps.

\begin{table*}
\centering
\begin{tabular}{cc}
\hline
\hline
MIPS sources & \# \\
\hline
\hline
MIPS sources with fl$_{24}$$\geq$82 $\mu$Jy  &  4417  \\
\hline
MIPS sources with a unique counterpart with R$\leq$24.1 & 2527 \\
MIPS sources with a non-unique counterpart with R$\leq$24.1 & 219 \\
MIPS sources without any counterpart with R$\leq$24.1 & 1671 \\
\hline
MIPS sources with a unique counterpart with R$\leq$24.1 and a reliable z$_{spec}$ & 1671 \\
\hline
\hline
\end{tabular}
\caption{Optical identification of the MIPS sources.}

\label{mips}
\end{table*} 

\subsection{Sample definition}
\label{Samp}
The flux limit of f$_{24}$=82 $\mu$Jy ensures that the sample is complete for LIRGs up to z$\sim$1,
which we impose as the upper redshift limit (see Figure~\ref{env1p1}, discussed in more detail i
n Subsection~\ref{LIR-SFR}).
We exclude  LIRGs at z$ < $0.7 to avoid contamination by lower redshift sources, which may have a 
different environment dependence than z$\sim$1 sources. 
In addition, EGS spectroscopy is designed to probe the distant universe and only a small number of 
LIRGs with secure spectroscopic redshifts are found at lower redshifts ($\sim$80 with z$\leq$0.7), 
which prevents us from building a good comparison sample of local LIRGs and ULIRGs. The final 
sample of LIRGs and ULIRGs is made of 314 sources with 0.7$\leq$z$\leq$1 and L$_{IR}\geq$10$^{11}$L$_{\odot}$.


\subsection{Completeness of the Sample}
At least two main issues affect the completeness of the sample: 1.) the spectroscopic redshift incompleteness; 
and 2.) the unknown redshift distribution of the 24$\mu$m sources with R$>$24.1. These two points 
are discussed below.
 
\subsubsection{Spectroscopic redshift incompleteness}
\label{zspec}
A part of the incompleteness for sources 
with R$ \leq $24.1 is due to the DEEP2 selection, which has only targeted a fraction of the sources 
with 18.5$ \leq $R$ \leq $24.1 (see section~\ref{deimos}). The probability for a source with 
18.5$ \leq$ R $\leq $24 to have been observed is estimated to be $\sim$ 0.63.
Another issue is the DEEP2 {\it{failures}}, which are the sources that were targeted 
but for which no secure spectroscopic redshift was obtained. These represent $\sim$30\% of 
the sample of targeted sources. As a consequence, the overall completeness of the DEEP2 
sample is $\sim$44\% over the 0.7-1.3 redshift range, assuming the failures are
uniformly distributed in z.

However LIRGs and ULIRGs are generally optically brighter than the field sources, have emission lines, 
and are relatively blue galaxies (according to the bimodality definition in Willmer et al., 2006). 
Some blue failures have also been targeted 
with LRIS-B and the great majority of them tend to have z$>$1.4 (Willmer et al., 2006, C. Steidel, 2004, 
private communication). Since we have restricted our sample to the 0.7$ \leq $z$ \leq $1 range, the completeness should 
be higher. In addition, the fraction of DEEP2 failures for sources with a similar R-I color as 
LIRGs and ULIRGs is close to 25\%, which is lower than the mean failure rate. 

To improve the completeness estimate of our sample, we use the photometric redshifts of CFHTLS. 
The CFHTLS only covers $\sim$70\% of the EGS field and we are only able to include 3163 of the 
4417 MIPS sources. The MIPS source and photo-z catalogs were matched with a 2$\arcsec$ search radius. 
Unique associations are found for 2187 24 $\mu$m sources\footnote{This fraction is similar to that 
found when matching the 24$\mu$m catalog to the DEEP2 photometric catalog.}, among which 1717 have a secure photo-z.

\begin{figure}
       \resizebox{\hsize}{!}{\includegraphics{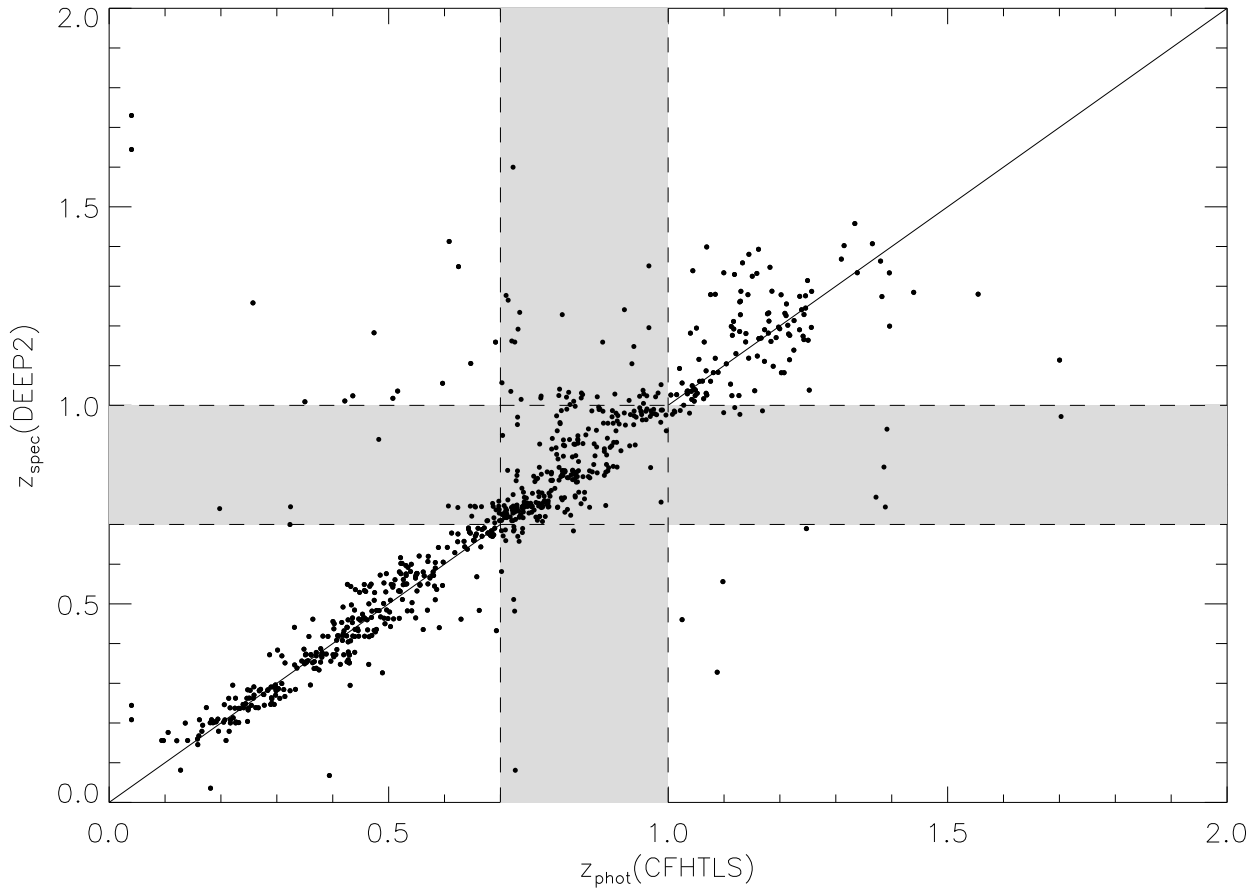}}
      \caption {Comparison between the photometric and spectroscopic redshifts from CFHTLS and DEEP2 respectively for sources detected at 24 $\mu$m above the 82 $\mu$Jy completeness limit. The dashed lines and grey area show the redshift range we focus on. Despite the photometric redshifts being less accurate for IR sources, the uncertainties are small enough to use them to estimate the completeness of the LIRG and ULIRG sample over 0.7 $\leq$ {\it{z}} $\leq$ 1.  }
\label{zz}
\end{figure}

Since photometric redshifts rely more on the spectral shape and the presence of the Balmer break, 
they are less reliable for objects with intense starbursts. Emission lines associated with 
starbursting galaxies and bright enough to
affect the photometry are not incorporated in an 
accurate way in the SED templates. IR sources are also affected by extinction. 
For all these reasons, we have tested the accuracy of 
the photometric redshifts of the IR sources.
This test is based on the 787 galaxies with DEEP2 spectroscopic redshifts with z$_q\geq$3
among the 24$\mu$m sources. The comparison of the two redshifts 
is presented in Tables~\ref{photoz} and ~\ref{zbin} and Figure~\ref{zz} for z$ < $0.7, 
0.7$ < $z$ < $1 and z$ > $1 bins. 
We find that the photometric redshift is typically 50\% less accurate for IR sources than for the whole 
sample and that the rate of catastrophic errors ($\eta$=$\sigma$$_{\Delta z/(1+z)}$$>$0.15) is twice as large.
These results are in agreement with those found for starbursting sources in Ilbert et al. (2006). 
Despite the photometric redshifts being less accurate for IR sources, the uncertainties are small enough 
to use them to estimate the completeness of the LIRG and ULIRG sample over 0.7 $\leq$ {\it{z}} $\leq$ 1; 
however, they will not be used for the environment study in the following sections. 
The sample 
of LIRGs defined in section~\ref{Samp} represents $\sim$57\% of the LIRGs with a photometric redshift 
in the same redshift range and with 18.5$ < $R$\leq$ 24.1. This completeness level is significantly 
higher than that for the full EGS DEEP2 sample. We have also used the photometric redshifts to estimate 
the completeness of the overall DEEP2 survey 
for 0.7$\leq$ {\it{z}} $\leq$1 and 18.5$\leq$ R $\leq$24.1, independent of infrared detections, and 
find 53\% of the sources 
with 0.7$\leq$ {\it{z}} $_{phot}<$1 and 18.5 $\leq$ R $\leq$24.1 have a spectroscopic redshift 
in the same range. The difference between the two completeness results (53 and 57\%) might be due to 
the uncertainties in  the photometric redshifts, $\sigma$$_{\Delta z/(1+z)}$=0.029, which becomes 
$\sigma$ $\sim$ 0.06 at z=1.

\begin{table*}
\centering
\begin{tabular}{ccc}
\hline
\hline
& $\sigma$$_{\Delta z/(1+z)}$  &  $\eta$=$\sigma$$_{\Delta z/(1+z)}$$>$0.15 \\    
\hline
IR sources         &         0.043                &    6.2\%                                    \\
whole sample$^1$   &         0.029                &    3.8\%                                     \\
\hline
\hline
\end{tabular}
\caption{Accuracy of the photometric redshifts for the LIRG and ULIRG sources compared to that for the whole sample of 
photometric redshifts for the CFHTLS.} 
\label{photoz}
\noindent{\footnotesize {\it Comments} : (1) from Ilbert et al. (2006)} 
\end{table*}

\begin{table*}
\centering
\begin{tabular}{cccc}
\hline
\hline
                & z$_{DEEP2}<$0.7 & 0.7$\leq$z$_{DEEP2}<$1 & 1$<$z$_{DEEP2}$  \\    
\hline
z$_{CFHTLS}$ $<$ 0.7         & 89 \% & 7 \% & 4 \% \\
0.7 $<$ {\it{z$_{CFHTLS}$}} $\leq$ 1     & 5 \% & 83 \% & 12 \%\\
1$<$z$_{CFHTLS}$             & 3 \% & 10 \% &86 \%\\
\hline
\hline
\end{tabular}
\caption{Accuracy of the CFHTLS photometric redshifts for LIRG and ULIRG sources as a function of redshift.} 
\label{zbin}
\end{table*} 
\subsubsection{Redshift range of the sources with R$>$24.1}
\label{sec:faint}
There are 1671 sources detected with MIPS at 24 $\mu$m that do not have counterparts with R $\le$ 24.1. 
These sources represent approximately 38\% of the 24$\mu$m sample and could significantly affect our 
conclusions if they fall in the 0.7 $\leq$ z $\leq$ 1 range. Sources that are optically faint and IR 
bright have been studied with IRS (Houck et al., 2005, 
Yan et al., 2007, Weedman et al. 2006). These extreme sources are mainly found 
at z$>$1. However, the sources in our sample are $\sim$ 10 times fainter at 24 $\mu$m than these high redshift sources. 

Since we cannot compute photometric redshifts based on the optical data for such faint sources, we used 
the IRAC colors to measure the wavelength of the so called ``stellar bump'' and estimate a rough fraction 
of sources with {\it{z}} $\leq$1. 
This broad feature is associated with the spectral peak at $\sim$ 1.6 $\mu$m caused by red giant and 
super-giant stars; it can provide a 
rough photometric redshift estimate if the stellar component dominates at the NIR wavelengths 
(Simpson \& Eisenhardt 1999, Le Floc'h et al., 2004). 
Weedman et al. (2006) use this bump in addition to the optical criterion m$_r$ $ > $23 to pre-select 
``starbursting sources'' at high redshift for IRS observation. The IRS spectra show that except for one object at z=0.98, 
all of the selected sources are at {\it{z}} $\sim$ 1.5-2. 

To estimate the redshifts, we first cross-correlated the optically faint MIPS sources with the IRAC data. 
There are 1260 MIPS sources with a unique counterpart out of the 1373 MIPS sources that are located 
in the IRAC field.
Among them, $\sim$ 41\% and 33\% of the sample tend to exhibit a ``stellar bump'' 
around $\sim$4.5 $\mu$m (channel 2) and $\sim$5.8$\mu$m (channel 3), respectively, giving a 
crude redshift range estimation of z$\sim$1.8 and $\sim$2.5. 15\% of the remaining sources 
exhibit a power law SED in the IRAC bands; such sources have been studied 
in the CDFS and tend to be at z$>$1 (Alonso-Herrero et al.; 2006; Donley et al.; 2007; 
Papovich et al.; 2007). 7\% show a stellar bump compatible with being at 3.6 $\mu$m or lower 
wavelength and could be 
sources at z$\sim$1.2 or below.
The rest of the sample cannot be classified. These sources exhibit a decrease in the IRAC SED 
from  3.6 to 5.8 $\mu$m and an increase at 8 $\mu$m.
This can be interpreted as sources for which the PAH emission band at 6.2$\mu$m is detected at 8 $\mu$m. 
If this interpretation is right, these sources could be at a z$ < $1. As a consequence, between 4 to 
11\% of the optically faint sources could be at z$ < $1.



\subsection{Is the ``LIRG and ULIRG'' sample unbiased?}

\begin{figure}
      \resizebox{\hsize}{!}{\includegraphics{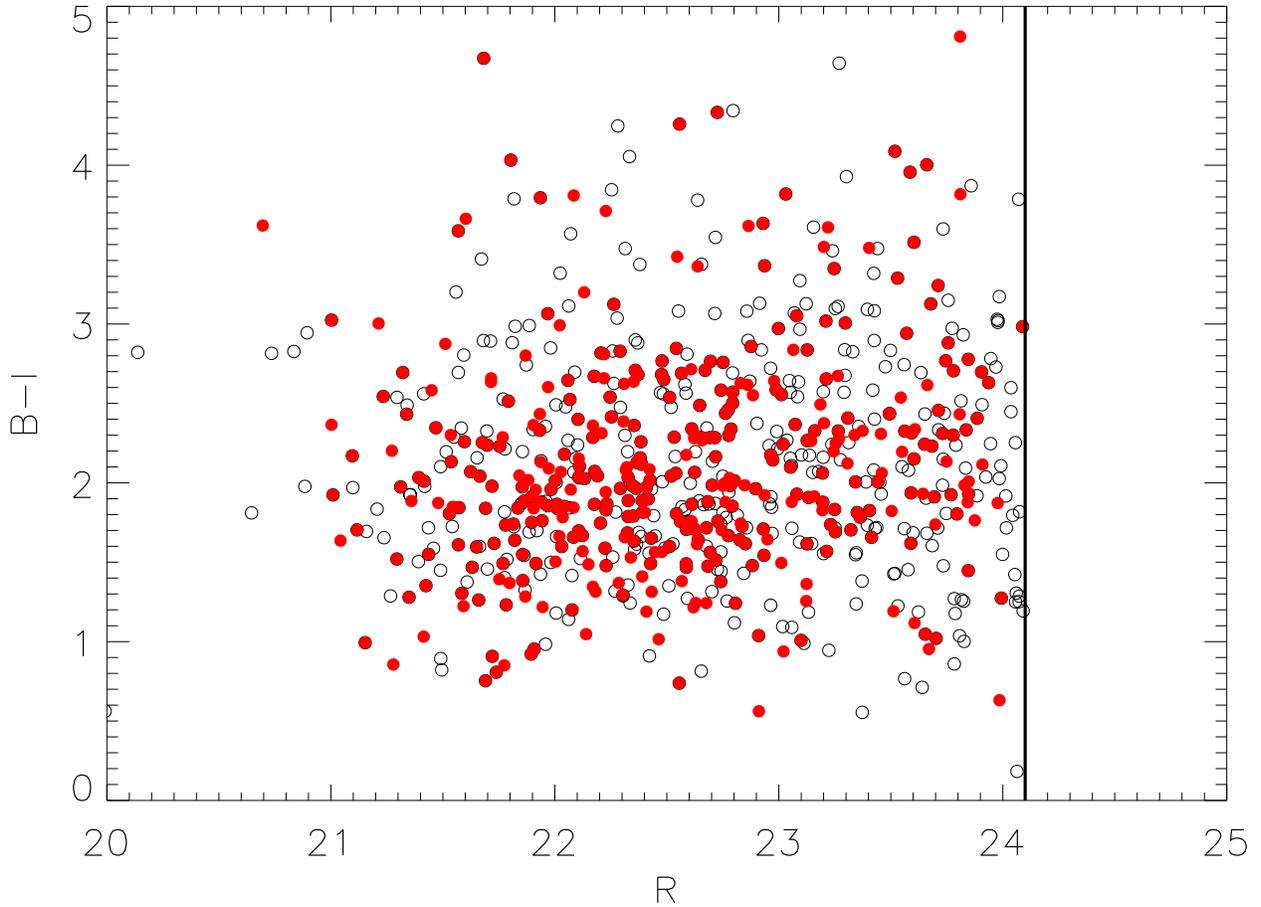}}
      \caption {B-I color as a function of R for IR--galaxies with 0.7$\leq$zphot$\leq$1.0 (open black circles) and IR--galaxies with  0.7 $\leq$zspec$\leq$1.0 (red solid circles) 
The vertical line is the R=24.1 DEEP2 magnitude limit. The plot shows that both samples cover a similar range in apparent magnitude and color suggesting there is no bias in the spectroscopic sample. [{\it{See the electronic edition of the Journal for a color version of this figure.}}]}
\label{BIAS}
\end{figure}

Since our LIRG and ULIRG sample is not complete, we need also to test how representative it is.
Due to the high density of sources that satisfy the DEEP2 criteria required for a source to be targeted, various
selection rules plus a final random selection were applied among multiple candidates to fill the 
DEIMOS slit masks (see Subsection~\ref{deimos}). None of these criteria should introduce a strong bias in 
the sample of IR sources.
To check, we compare the color-magnitude distribution in Figure~\ref{BIAS} for the sources with 
18.5$ < $R$ < $24.1 for which the CFHTLS 
gives 0.7$\leq$z$_{phot}<$1 (open black circles) to the spectroscopic sample of LIRGs and ULIRGs. 
Figure~\ref{BIAS} shows that the spectroscopic sample of galaxies used in this work covers the same 
parameter space as the larger DEEP2 parent sample.

\section {AGN contamination}

IR galaxies are known to be powered by both AGN and star formation activity. It is not 
known whether galaxies dominated by AGN will lie in similar 
environments as star-forming ones. We use X-ray and IRAC data to isolate X-ray detected AGN and IRAC power-law ones to estimate the AGN contamination of our sample.

\subsection{MIPS sources with X-ray emission}
\begin{figure}
       \resizebox{\hsize}{!}{\includegraphics{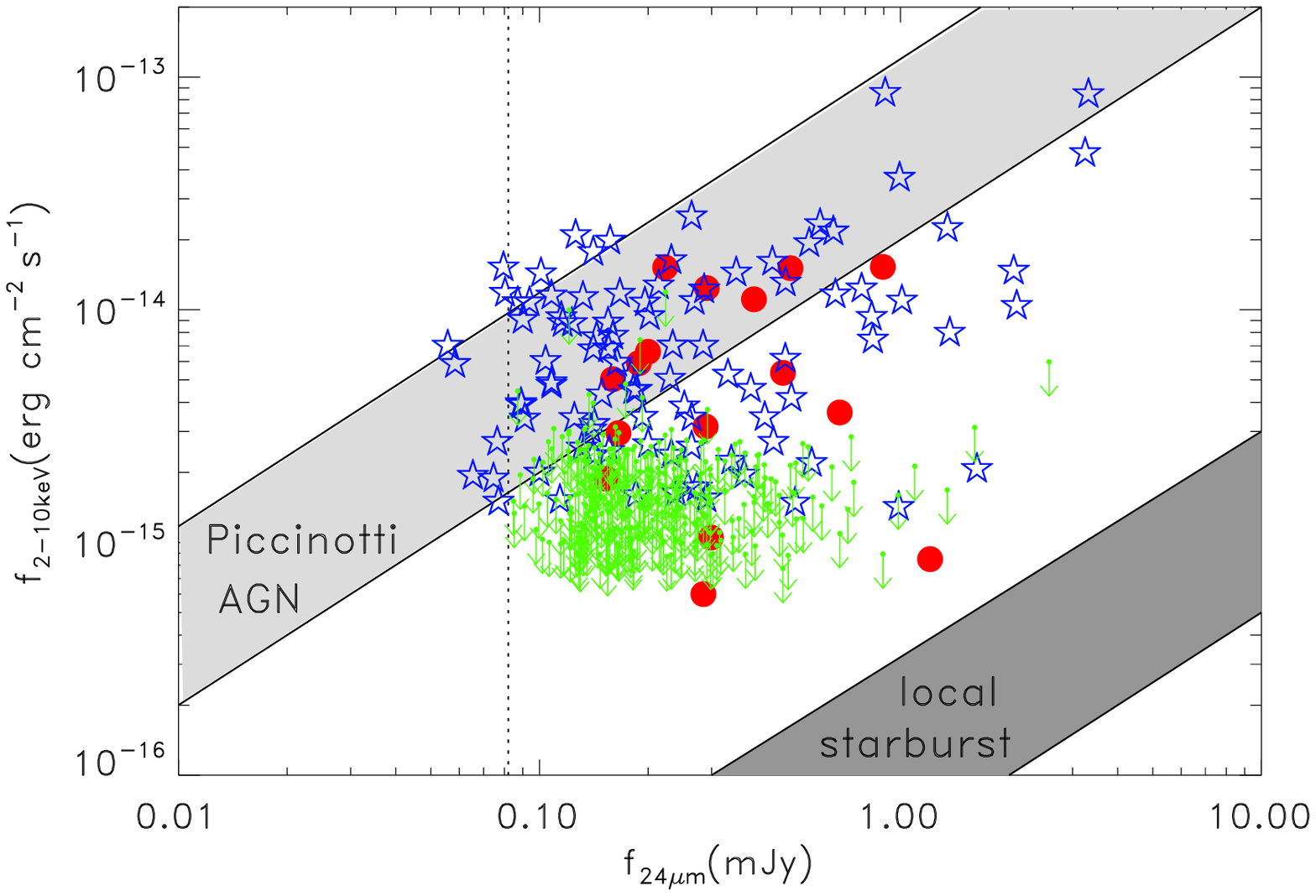}}
      \caption {f$_{24}$ as a function of the X-ray flux in the hard band. {\bf{Solid red circles :}} Sources 
detected in the hard X-ray band. {\bf{Open blue stars }} : X-ray sources detected in the 1 M sec field in the CDFS with a 24 $\mu$m counterpart (Rigby et al. (2004). {\bf{Green arrows :}} IR--galaxies 
with an X-ray upper limit. The lightly 
shaded area is the extrapolation of the median X-ray to mid IR ratio of local hard X-ray selected AGN (Piccinotti 
et al., 1982) while the dark shaded area is an extrapolation of the local starburst relation from Ranalli et al. (2003). 
The extrapolation is described in Alonso-Herrero et al. (2006). The dashed line is the MIPS completeness limit. This figure shows that the AGN contamination is weak as seen from the distribution of upper limits represented by the green arrows. [{\it{See the electronic edition of the Journal for a color version of this figure.}}]}
\label{X}
\end{figure}

X-ray data are a primary way to identify AGN activity.
We matched the X-ray catalog with the optical counterparts of the LIRGs because the astrometry 
is better for the optical. The PSF of Chandra is not uniform over the field of view, varying  
from $\sim$0.5 $\arcsec$ (FWHM) at the center, increasing up to 4 $\arcsec$ (FWHM) close to the edge.  
The search radius for the X-ray optical correlation has been changed adaptively as a function 
of the off-axis angle to account for this effect.
Figure~\ref{X} presents the 24 $\mu$m flux versus the 2-10 keV X-ray flux of the LIRGs detected 
in the X-ray hard band (red solid circles) or the upper limits of undetected sources (green arrows). 
The lightly shaded area represents the extrapolation of the median X-ray to mid IR ratio of local hard X-ray selected 
AGN (Piccinotti et al., 1982) while the dark shaded area is an extrapolation of the local starburst 
relation from Ranalli et al. (2003)\footnote{See Alonso-Herrero et al., 2004 for more details}. 
A total of 20 LIRGs are detected with at least one measurement in one of the X ray bands (17 in the soft band, 16 in the hard band and 14 in the ultra hard one), which leads to a contamination of $\sim$6\% of the LIRG and ULIRG sample. Twenty other sources have upper limits falling within the  Piccinotti et al. (1982) region and could be AGN, 
which gives an upper limit of $\sim$12\% for the AGN contamination.

One could argue that the X-ray data in the EGS are not deep enough and that as a result we miss some dusty AGN. 
Therefore, we also show the distribution of the hard X-ray sources with MIPS counterparts at 24 $\mu$m in the CDFS 
(Rigby et al., 2004, blue stars). Most of the AGN from Rigby et al. (2004) are X-ray brighter than the majority 
of the X-ray upper limits of the LIRGs studied here indicating that we must have missed relatively few AGN 
when compared to the deepest X-ray fields.

\subsection{IRAC color selection}

The IRAC colors have been used extensively to find dusty active galaxies not detected in the X-ray 
(Lacy et al., 2004, Stern et al., 2005, Alonso-Herrero et al., 2006, Barmby et al., 2006, Donley et al., 2007).
An object 
dominated by a dusty AGN will show little sign of the stellar bump and its IRAC SED will be fitted by a red 
power law with f$_{\nu}$ $\propto$ $\nu^{\alpha}$ where $\alpha < $0. The range of $\alpha$ allowed to define 
power-law AGN varies and can be defined as $\alpha$$ < $-0.5 (Donley et al., 2007, Alonso-Herrero, 2006) 
or $\alpha \leq$0 (Barmby et al., 2006). One of the reasons to choose a stricter 
definition (e.g., $\alpha \leq$-0.6) for the power law 
sources is that the redshift range of these sources is usually unknown and the goal is to ensure the sources 
selected are really dominated by a power-law AGN. In our case, the redshift range of the sources is known 
and we can reasonably expect sources with an IRAC SED fitted with -0.5$\leq$$\alpha$$ < $0 to be dominated 
or partly powered by an AGN. We also want to estimate an upper limit on the contamination by 
the IRAC power-law sources in our sample, so we will consider both ranges of $\alpha$. Five sources are found 
with $\alpha\leq$-0.6, while 11 are found with  $\alpha\leq$0 among the 283 LIRG sources with 
a detection in all the IRAC bands. 
The small number of 
power-law sources detected in this sample is not very surprising since a large fraction of the power-law 
sources is found at higher redshift ( Alonso-Herrero, 2006, Donley et al., 2007, Papovich et al., 2007).
Among these 11 sources, 5 are X-ray detected in the hard band, which shows that these techniques of selecting AGN are complementary.

\subsection{AGN contamination among the LIRGs and ULIRGs}

At low redshift, La Franca et al (2004) show that at least 25\% of the MIR sources detected in the 
ELAIS-S1 field are associated with an AGN according to their optical spectra; they stress that this percentage 
is a lower limit since some objects whose spectra do not show signs of AGN activity are detected in the X-ray. 
However, the presence of an AGN does not imply that it dominates 
the total infrared luminosity. Using SED modeling, Rowan-Robinson (2005) showed that the total infrared 
luminosity is dominated by AGN activity for only 11\% of the total ELAIS sample. 

In the redshift range we are studying, Fadda et al. (2002) combine X-ray with MIR 15$\mu$m data in the Lockman 
Hole and HDFN to conclude that the AGN contribution is about 17\% of the 15 $\mu$m 
background. Liang et al. (2004) studied the optical emission lines 
and conclude that about 23\% of the LIRGs detected at 15 $\mu$m are associated with AGN. Weiner et al.(2007) 
studied the AGN contamination of all the DEEP2 galaxies in the 0.35$ < $z$ < $0.81 redshift range in the 
pseudo-BPT diagram (independent of the MIR detections) and find that 2-3\% of the blue galaxies with 
emission lines could still be dominated by AGN, while 1/3 of the red galaxies with emission lines are close to 
the star-forming track, where blue and red galaxies are defined in Willmer et al. (2006). 
Combining these results for LIRGs and ULIRGs, we find that $\sim$13\% of the sample could be contaminated by AGN. However, since the red LIRGs are bluer than most of the red sequence, this estimate of AGN contamination may be high.

However, even if AGN activity is detected, the star-formation activity can still dominate the IR luminosity 
of these sources (Oliver \& Pozzi, 2005, Rowan-Robinson, 2005, Le Floc'h et al., 2007). Because of the 
uncertainty in the main engine of the IR emission, we have not removed these sources from the sample, 
but we have checked that their removal does not change our conclusions.
\label{env}

\section{Quantifying the local environment of LIRGs and ULIRGs.}
\label{envvv}
The goal of this section is to probe whether LIRGs at {\it{z}} $\sim$ 1 are found
preferentially in over-dense or under-dense environments. We use the local environment 
indicator $\delta_3$ defined in section~\ref{delta3} 
and refer to it as the quantity D=log$_{10}$(1+$\delta_3$).

\subsection{SFR(IR)--local density relation.}
\label{LIR-SFR}
The total infrared luminosity (L$_{IR}$) is derived using the 24 $\mu$m data.
The continuum of the mid-infrared emission of galaxies is dominated by the output 
of the very small grains (VSG, Desert, Boulanger \& Puget, 1990). To this continuum must be added 
broad aromatic emission features at  6.2, 7.7, 8.6, 11.3 and 12.7\,$\mu$m (PAH, Puget \& Leger, 1989). 
At z$ < $0.6 the 24 $\mu$m emission probes the VSG continuum, which has been shown to 
correlate very well with L$_{IR}$ (Chary \& Elbaz, 2001, Alonso-Herrero, 2006). 
At higher redshifts and up to z$\sim$1.2, the 12 $\mu$m PAH complex enters the MIPS 24 $\mu$m 
band. The luminosity at 12 $\mu$m has also been shown to correlate well with the L$_{IR}$ in the 
local universe (Chary \& Elbaz, 2001) and up to z$\sim$1.2 (Marcillac et al., 2006).

We estimate the infrared luminosity of each galaxy using the spectral energy 
distribution (SED) from the Dale \& Helou (2002) library as explained in Marcillac et al. (2006). 
Each of the template SEDs is redshifted to match the galaxy. 
Then, the SED with MIR flux density closest to the observed value is used to 
derive L$_{IR}$ after it is normalized 
to the exact observed flux density. Marcillac et al. (2006) compared L$_{IR}$ estimated using 24$\mu$m 
data to L$_{IR}$ estimated using radio data for LIRGs and found a scatter of 40\%  between 
the two estimates. 
Using stacking analysis (Papovich et al., 2007) 
or direct FIR detection (Frayer et al., 2006) 
it appears that the FIR emission of distant IR sources is in agreement with SED 
templates, which are calibrated on local IR sources, up to z$\sim$1.

\begin{figure}
      \resizebox{\hsize}{!}{\includegraphics{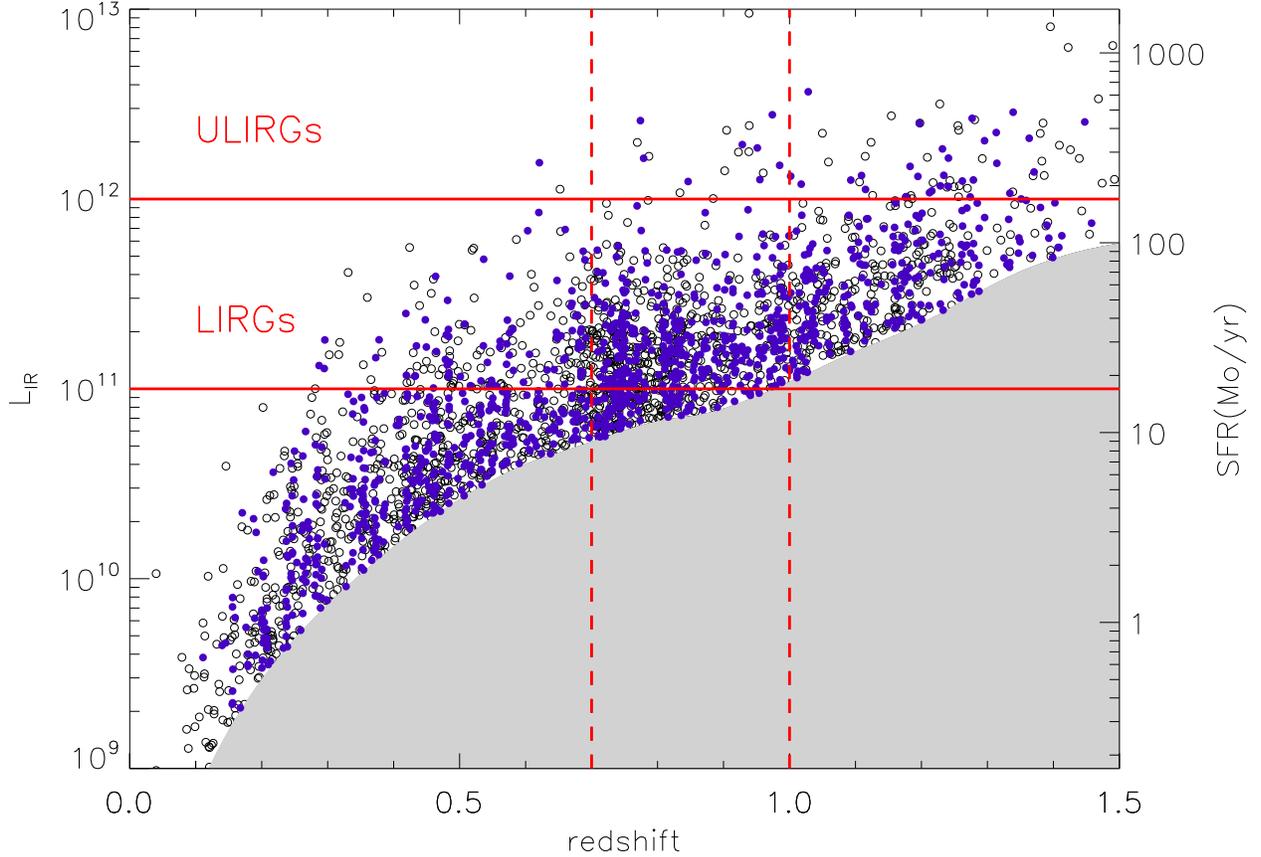}}
      \caption {Total infrared luminosity as a function of redshift for sources above the MIPS completeness 
limit of 82 $\mu$Jy (grey region). {\bf{Solid blue circles :}} sources with a secure spectroscopic redshift in DEEP2. 
{\bf{Empty circles :}} sources with photometric redshift. The red dashed lines show the redshift range where we focus. The star formation rates in units of solar masses per year, SFR(IR) were derived using the 
relation of Kennicutt (1998) :
SFR(IR,M$_{\odot}$/yr)=1.71 $\times$10$^{-10}$(L$_{IR}$/L$_{\odot}$). {\bf{Red solid lines :}} show the LIRG and ULIRG regimes. }
\label{LIRR}
\end{figure}

\begin{figure}
       \resizebox{\hsize}{!}{\includegraphics{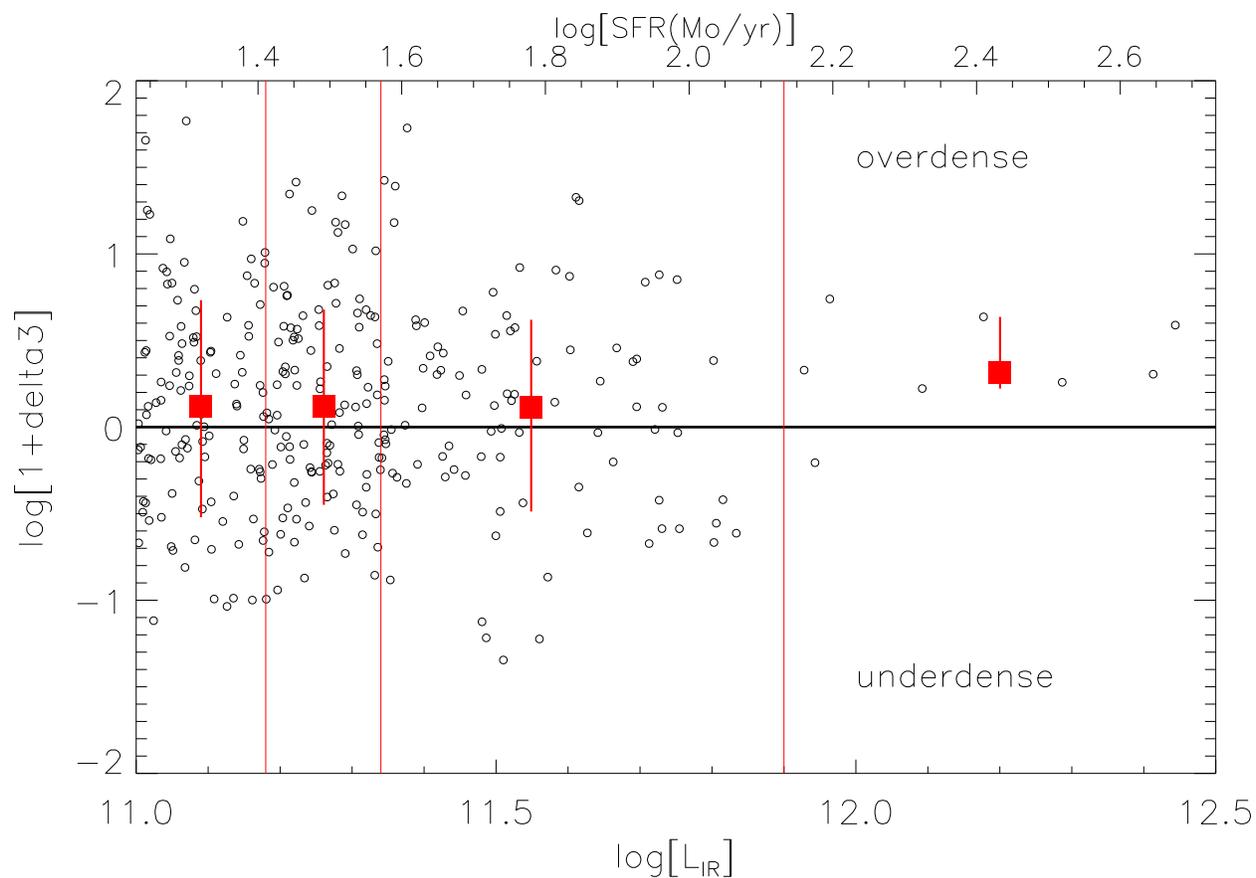}}
      \caption {Local environment of LIRGs as a function of L$_{IR}$. The data have been divided into four 
bins : the first three are designed to approximately contain the same number of objects ($\sim$100), while 
the latter is made of ULIRGs and the brightest LIRGs. The red squares are the median values in each bin and 
the error bars are a 1-$\sigma$ dispersion estimated on each side of the median value. [{\it{See the 
electronic edition of the Journal for a color version of this figure.}}]} 
\label{c}
\end{figure}

Figure~\ref{LIRR} shows the evolution of L$_{IR}$ with z; The grey area shows the trend 
of the completeness limit of the IR luminosity with redshift. 
For LIRGs and ULIRGs, the star formation is dominated by the infrared emission (Bell et al., 2005) 
and we use the relation of Kennicutt (1998) to estimate the SFR from L$_{IR}$.
scenarios and galaxy evolution has often been discussed for local and distant LIRGs and ULIRGs 
(Hammer et al.,2005, Sanders et al., 1988).
Figure~\ref{c} presents the local environment of the IR galaxies as a function of L$_{IR}$. We have 
divided this figure into 4 luminosity bins: the first three are designed to have approximately the 
same number of objects ($\sim$100), while the last is made of ULIRGs and the brightest LIRGs. The red 
squares are the median values in each bin and the error bars are a 1-$\sigma$ dispersion estimated 
on each side of the median value. The figure shows that there is no strong dependence between L$_{IR}$ 
and the local density up to L$_{IR}$ $\sim$ 7-8$\times$10$^{11}$L$_{\odot}$.
This result suggests that once 
a galaxy becomes a LIRG or a ULIRG, it is unlikely that the environment has a significant influence
on the SFR activity in a substantial number of sources, at least in or near the dense 
environments that are probed in the EGS. This is in accordance with results found in the local 
universe where the star-forming sources (with EW(H$\alpha$)$>$4\AA$\,$) do not show any correlation 
between EW and local environment density (Balogh et al., 2004). 
scale to prevent us from finding a trend. For example mergers of galaxies could help consume the gas 
very quickly, feeding both an AGN and an intense burst of star formation. However, only a small 
fraction of LIRGs and ULIRGs at 0.7$ < $z$ < $1 are associated with direct mergers and it is now 
accepted that the majority of LIRGs exhibit only slightly disturbed morphology, which tend to rule out this hypothesis.

At log$_{10}$(L$_{IR}$)$\geq$ 11.9, there is only one object in an under-dense environment 
(log$_{10}$(1+$\delta3) < $0), whereas 8 sources lie in over-dense environments. One of these 
has an AGN since it is detected in all the X-ray bands and another is classified as an IRAC 
power law object. The other sources do not have any AGN signature in the X-Ray or IRAC but a 
dusty AGN can not be excluded. Therefore, ULIRGs tend to be found in over-dense environments; 
this conclusion is weakened by the small number of ULIRGs in this field.
We used Monte Carlo simulations to estimate the significance of this tendency. We use the whole 
sample of LIRGs (without ULIRGs) and find a probability of $\sim$15\% to have 5 of 6 sources in overdense environments. 
This result is in agreement with Lin et al. (2007) who find a mid-IR 
enhancement in pairs separated by less than a few tens of kpc (Lin et al. 2007) for ULIRGs. 

However, there is another possible explanation since the sources with log$_{10}$(L$_{IR}$)$\geq$ 11.9 are 
also optically brighter (M$_{B}$=-21.43$_{-0.56}^{+0.48}$) 
than the ones fainter in the IR (M$_{B}$=-20.64$_{-0.53}^{+0.71}$). 
The optically brighter sources are also in denser environments at z$\sim$1 (Cooper et al., 2007). As a 
consequence, the fact that the local environment of ULIRGs tends to be denser can be interpreted with 
the density - luminosity relation alone since ULIRGs tend to have a similar environment to sources without 
IR detection and similar optical luminosity. 

\subsection{Comparing the local environment of LIRGs and ULIRGs with that of 0.7 $\leq$ z $\leq$ 1 optically (R$\leq$24.1) selected galaxies}
\label{toto}
\begin{center}
\begin{figure*}
       \resizebox{14cm}{!}{\includegraphics{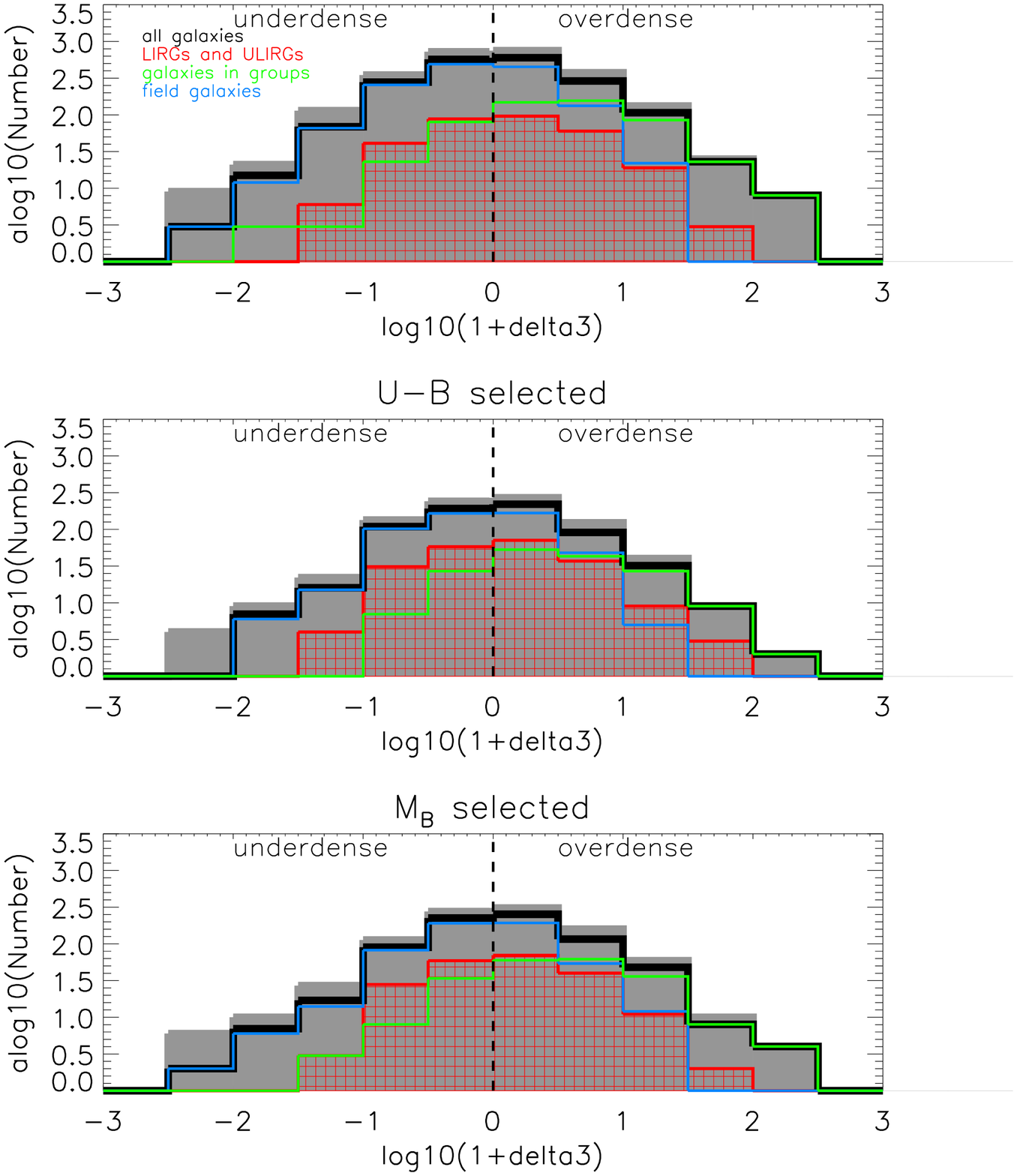}}

      \caption {Comparison of the local environment of the LIRG and ULIRG sample (red) with the ``all galaxy'' sample 
(grey area). The sources classified in Gerke et al. (2007, black) have been divided into two categories : the 
``group'' sample, which is composed of sources belonging to groups with Vdisp $>$100 km $\,$s$^{-1}$ (green), 
and field sources (blue). {\bf{Top}} : The samples are composed of all the sources with 0.7$ < $z$ < $1. {\bf{Middle : }}The 
samples are composed of sources with 0.7  $\leq$ z $\leq $1 and 0.70 $\leq$ U-B $\leq$ 1.04 which is the color range of the LIRGs and ULIRGs. {\bf{Down:}} The samples are composed of sources with 0.7 $< $z$ < $1 and -21.18 $\leq$ M$_B$ $\leq$ -19.93 which is the absolute magnitude range of the LIRGs and ULIRGs.}
\label{env1p1}
\end{figure*}
\end{center}

\begin{center}
\begin{figure*}
       \resizebox{14cm}{!}{\includegraphics{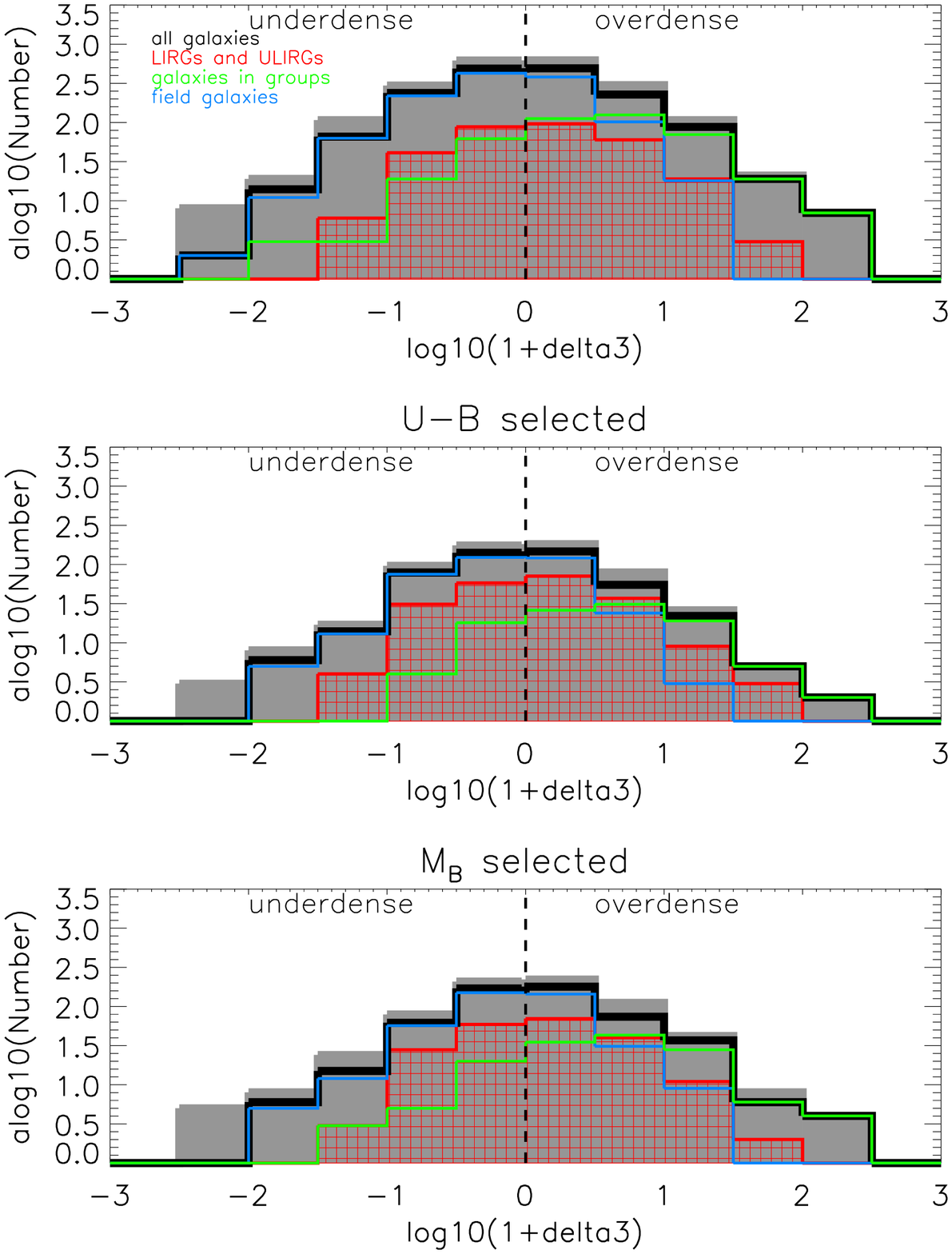}}       
\caption {Distributions as in Figure~\ref{env1p1} but now restricting the comparison samples to contain only non-IR galaxies.}
\label{LIR-IR}
\end{figure*}
\end{center}

We therefore compare 
the environment of LIRGs and ULIRGs to that of non-IR-emitting optically-selected galaxies at similar redshift. 
The distribution of D=log(1+$\delta_3$) for the ``LIRG and ULIRG'' sample is presented in Figure ~\ref{env1p1} (red histogram). We find  
that LIRGs and ULIRGs tend to be in over-dense environments, since $<$D$_{LIRG \& ULIRG}$$>$=0.12, which is also confirmed by the fact 
that a majority of LIRGs and ULIRGs, i.e. 57\% of them, are located in overdense regions (D$>$0).

 We have compared the local environment distribution for LIRGs and ULIRGs to that of a control sample with secure spectroscopic redshifts regardless of 24$\mu$m properties but located in the same redshift range (hereafter the ``all galaxy'' sample grey shaded area in Figure~\ref{env1p1}). Our findings are :
\begin{itemize}
\item The median LIRG and ULIRG environment is 1.2 time denser 
compared to the median environment of the ``all galaxy'' sample. 

\item LIRGs and ULIRGs tend to avoid very 
under-dense environments (D$ < $-1), since only 4\% of them have D$<$-1, which is 3 times less numerous than for the ``all galaxy'' sample.
This suggests that a minimum density may be needed to trigger IR activity. 

\item A Kolmogorov-Smirnov (KS) test confirms that the distribution of the local environments of the ``LIRG \& ULIRG'' and ``all galaxy'' samples  differ significantly since the probability is $\sim$95\%
 that the distributions come from  
different parent populations. 
Once we remove the sources detected above the MIPS completeness limit belonging to the ``all galaxy'' sample (``all galaxy without IR'' sample, hereafter), the KS test shows even stronger evidence that the local environment distribution of LIRGs and ULIRGs and of the ``all galaxy'' sample come from different populations
. This comparison is shown in Figure~\ref{LIR-IR}a, where it can be seen that LIRGs and ULIRGs reside in significantly denser environments compared to other sources detected in the 
same redshift range.
 \end{itemize}
This confirms the previous results of Cohen et al. (2000) who find that IR galaxies in the HDFN tend to belong to redshift peaks. This is also in agreement with Elbaz et al. (2007) and Cooper et al. (2007) who find that the 
mean SFR correlates with density at z$\sim$1.

We have also compared the local environment of LIRGs and ULIRGs to that of sources at similar redshift 
and classified as group members (the ``group sample'', green histogram hereafter) and field members 
(the ``field'' sample, blue histogram hereafter) in Gerke et al. (2007). 
The D distributions of all these samples are presented in Figure~\ref{env1p1}a. The black line indicates 
the D distribution of the sources that have been classified in Gerke et al. (2007)\footnote{They did not 
classify all the sources in the EGS spectroscopic redshift catalog because the EGS sampling 
is higher than in the three other fields studied in DEEP2. Since the group finding algorithm is 
optimized for a specific sampling, they re-sampled the EGS field data. 
It must be noticed that if the black histogram is used instead of the ``all galaxy sample'', the previous 
results remain unchanged, which emphasizes that the two samples have very similar properties.}.
Our main findings are : 

\begin{itemize}
\item The local environment of LIRGs and ULIRGs is intermediate between the ``field'' and ``group'' sources where $<$D$_{field}$$>$=-0.11 and $<$D$_{group}$$>$=0.52.

\item The lowest value of D for LIRGs and ULIRGs is similar to the lowest value of D for the ``group sample'' , while the highest is close to 
the ``field sample'' one. This implies that the LIRGs and ULIRGs tend to avoid both very under-dense and over-dense 
environments compared to field and group sources.  
These results do not change significantly if the IR sources are removed from the ``all galaxy'',''field'' 
and ``group sample'' as is shown in Figure~\ref{LIR-IR}a. 
\end{itemize}
These two points will be discussed in section~\ref{env_conc}.

\subsection{ LIRGs and ULIRGs and the color-environment relation.}
\label{colsub}

We discussed in the introduction how the environment of local galaxies correlates with 
galaxy properties such as the rest-frame colors and magnitude. Some of these relations have 
been extended up to z$\sim$1 (Cooper et al., 2006). The goal of this subsection and the following 
one is to study the environment of LIRGs and ULIRGs in the framework of the color-environment and 
luminosity-environment relations.

Cooper et al. (2006) find that 
the fraction of red galaxies (i.e. galaxies on the red sequence, see Willmer et al., 2006) 
depends strongly on environment at least up to z$>$1. We compare the mean local environment of LIRGs and 
ULIRGs to the one of both blue and red galaxies : In both cases, the KS test shows a significance level 
of 99.1\% that the parent populations are different. The red galaxies exhibit a local mean environment 
that is $\sim$1.9 times denser than the blue galaxy one; this value is slightly higher than the one 
given in Cooper et al. (2006) because we do not use the same samples of sources. However, the 
mean local environment of the red galaxies is only 1.3 times denser than the LIRG and ULIRG one. That 
is, the LIRGs and ULIRGs have a mean local environment that is intermediate between that of the blue and red galaxies.

\begin{center}
\begin{figure*}
       \resizebox{12cm}{!}{\includegraphics{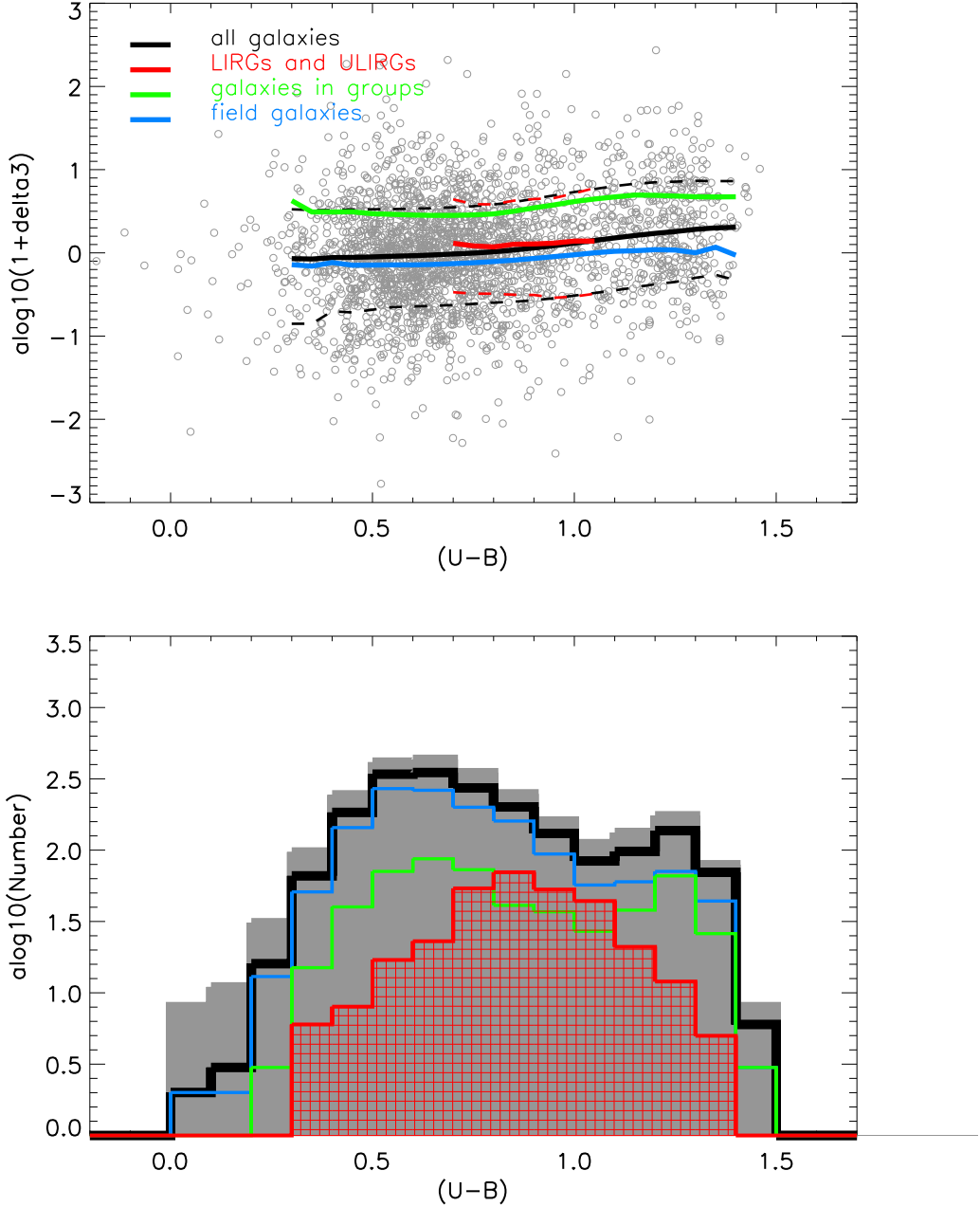}}
      \caption{{\bf{Top :}} Environment as a function of the (U-B) color. The black, red, green and blue lines show the mean evolution of the ``all galaxy'',''LIRG and ULIRG'',''group'' and ``field'' samples. 
They have been computed in sliding boxes with $\Delta$(U-B)=0.15 for the four subsamples.The dashed lines show the 1$\sigma$ error bars estimated in a sliding box with similar size. Due to the lack of LIRGs and ULIRGs outside the ``green valley'', the mean trends have been represented for 0.70$\leq$(U-B)$\leq$1.07 which is the color range of the 68\% of the LIRGs and ULIRGs. The mean local environment of LIRGs and ULIRGs is very similar to the one of sources with similar colors, but still remains denser than that of the ``field'' and less dense than the ``group'' sample.   
{\bf{Bottom :}} Comparison  of the histograms of (U-B) colors for all the ``all galaxy'',''LIRG and ULIRG'',''group'' and ``field'' samples.  }
\label{UB}
\end{figure*}
\end{center}

We now compare the local environment of LIRGs and ULIRGs to the one of sources with similar optical 
colors : Figure ~\ref{UB}a reproduces Figure 5 of Cooper et al. (2006) but restricting the sample 
to 0.7$ \leq $z$ \leq $1 in the EGS field. It shows the trend of the local environment with respect to (U-B) for the 
``all galaxies'', ``field'', ``group'' and ``LIRG \& ULIRG'' samples. The thick green, blue, 
red\footnote{Due to the lack of LIRGs and ULIRGs outside the ``green valley'', the mean trend and 
dispersion have been represented only on the 0.70$\leq$(U-B)$\leq$1.07 for LIRGs and ULIRGs on 
Figure ~\ref{UB}a.} and black lines\footnote{estimated using 
sliding boxes with $\Delta$(U-B)=0.15 for the four subsamples.} represent the running median 
value of D as a function of (U-B). 
The dashed lines show the 1$\sigma$ error bars estimated in the same sliding box for the ``LIRG \& ULIRG'' and ``all galaxy'' samples. 
The four samples show a strong dependence of D on (U-B) and do not have identical (U-B) distributions as shown in Figure ~\ref{UB}b. While the ``field'', ``group'' and ``all galaxy'' samples clearly show evidence of the color bimodality, the ``LIRG \& ULIRG'' one exhibits a single peak in the color corresponding to ``green valley''. This peak lies on the blue side of the bimodality with 84\% of the LIRGs and ULIRGs on the blue side of the color-magnitude relation. 
The first three samples exhibit a mean color of (U-B)=0.70$_{-0.20}^{+0.39}$, 0.76$_{-0.23}^{+0.45}$, 0.69$_{-0.18}^{+0.32}$ respectively whereas the LIRGs and ULIRGs exhibit a mean redder color with (U-B)=0.87$_{-0.17}^{+0.20}$.
As a consequence, the ``field'', ``group'' and ``all galaxy'' samples studied previously are biased toward 
sources with bluer colors and in under-dense environments.
Figure~\ref{env1p1}b presents the distribution of the ``LIRG'', 'group'', ``field'' and ``all galaxy'' 
samples restricted to the LIRG and ULIRG color range.
When the ``all galaxy'' sample is restricted to sources with the same (U-B) colors\footnote{The (U-B) range is 0.70$\leq$(U-B)$\leq$1.04 and has been chosen to include the median 
(U-B) with its 1$\sigma$ error bar. 68\% of the LIRGs and ULIRGs fall in this range.} as the LIRGs and ULIRGs, we find that the KS probability of these samples coming from the same parent population increases to $\sim$ 30 \%.

This result can be 
partly explained since all three samples are contaminated by LIRGs and ULIRGs. If we remove all IR sources from the ``all galaxy'', ``group'' and ``field'' samples (see Figure ~\ref{LIR-IR}b), 
the LIRGs and ULIRGs still exhibit a denser local environment than the ``all galaxy sample without IR sources''since a KS test shows a probability of 3\% that the ``LIRG \& ULIRG '' and ``all galaxy without IR'' local environment distribution come from the same parent population. However, this color cut might not be entirely satisfactory since some residual color trends might still remain in the subsamples and could bias the result. To test this possibility, we use the ``all galaxy sample'' to generate $\sim$ 100 subsamples with the same number of sources as 
the ``LIRG and ULIRG '' one and with a similar color distribution as the LIRGs and ULIRGs. Comparing the environment distribution of the LIRGs and ULIRGs with the Monte-Carlo simulations, we find that KS = 0.41$_{-0.25}^{+0.38}$ that both come from the same parent population. We also used the ``all galaxy without IR'' sample in a similar way and obtained 
$KS$ =0.14$_{-0.10}^{+0.26}$. These KS tests are in good agreement with the previous results.

The mean local environment of LIRGs and ULIRGs still remains denser than that of the ``field'' and less dense than the ``group'' sample. This trend remains valid with a color selection as well as for the ``all galaxy'' and ``all galaxy without IR'' samples . Thus, LIRGs and ULIRGs are not peculiar sources; they tend to obey the same relation between over-density and rest-frame (U-B) color as is followed by other galaxies at z$\sim$1.

However, it must be noted that even if the ``all galaxy'' and ``all galaxy without IR sources'' samples have similar M$_B$ 
( M$_{B,all galaxy}$=-20.08$_
{-0.76}^{+0.71}$ and M$_{B,all galaxy w/o IR}$=-19.89$_{-0.75}^{+0.66}$), the ``LIRG \& ULIRG'' sample is 0.5 mag more luminous 
( M$_{B,LIRGs and ULIRGs}$=-20.60$_{-0.58}^{+0.63}$). 
This shows that removing IR sources from the sample will bias the sample against more luminous sources with slightly different colors (see Fig.3 from Cooper et al., 2006).

\subsection{LIRGs and ULIRGs and the M$_B$-environment relation.}
\label{MBsub}
\begin{center}
\begin{figure*}
       \resizebox{13cm}{!}{\includegraphics{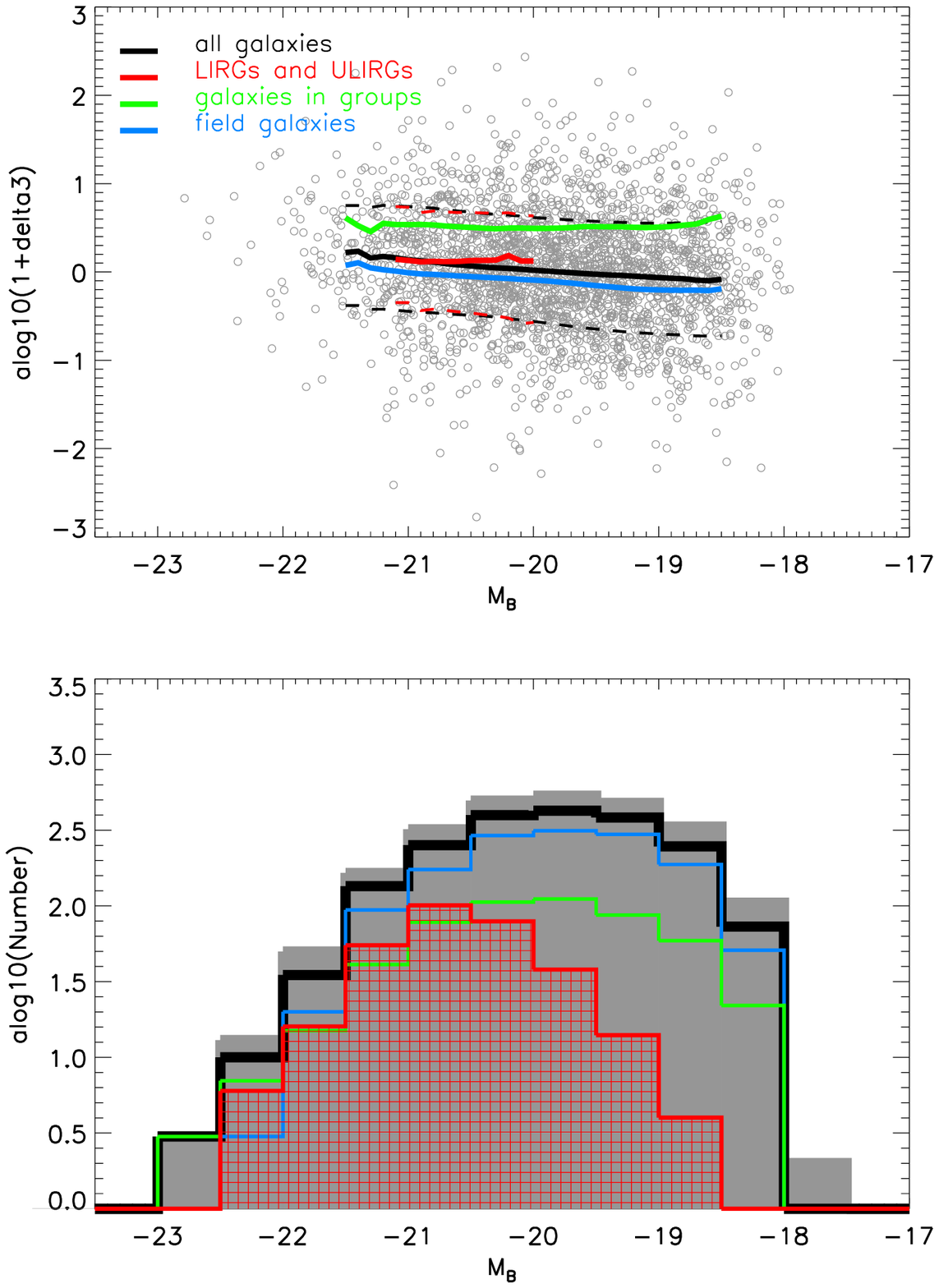}}
      \caption { {\bf{Top :}} Environment as a function of M$_B$. The black, red, green and 
blue lines show the mean trend of the ``all galaxy'',''LIRG'',''group'' and ``field'' samples. They 
have been computed in sliding boxes with $\Delta$(M)$_B$=0.3 for the four subsamples. The dashed lines show the 1$\sigma$ error bars estimated in a sliding box with similar size. The mean local environment of LIRGs and ULIRGs is very similar to the one of sources with similar optical luminosity, but still remains denser than that of the ``field'' and less dense than the ``group'' sample. {\bf{Bottom :}} Comparison of the histograms of the M$_B$ colors for the ``all galaxy'',''LIRG and ULIRG'',''group'' and ``field'' samples.  }
\label{MB}
\end{figure*}
\end{center}

Cooper et al. (2006) study the luminosity environment relation up to z$\sim$1. They show that  the luminosity is a function of environment for both blue and red galaxies at z$\sim$1 (see their Figure 9). 
They find that bright blue galaxies are in over-dense regions, contrasting with the local universe where no correlation has been identified between luminosity and over-density for blue galaxies. 
Cooper et al. (2007) interpret this relation as evidence for the different relation between SFR and environment at z$\sim$1. Figure ~\ref{MB}a shows the dependence of the local environment with respect to M$_B$ for the 
``all galaxies'', ``field'', ``group'' and ``LIRG \& ULIRG'' samples, where it is seen that bright sources tend 
to be found in over-dense environments. However, as is shown in Figure~\ref{MB}b, the LIRGs and ULIRGs have a median of 
M$_{B}$=-20.64$_{-0.53}^{+0.71}$, which is roughly one magnitude brighter than the mean luminosity of other sources located in the same redshift range. LIRGs and ULIRGs tend to occupy a very similar environment as the bright sources in the ``all galaxy'' sample. 

To quantify this result, we restricted all the samples to sources with similar M$_{B}$\footnote{-21.18$\leq$M$_B$$\leq$-19.93 (which contains 68\% of the LIRGs and ULIRGs, including the median value and 1$\sigma$ error bar.)} as the 
LIRGs and ULIRGs. The results are presented in Figure~\ref{env1p1}c (and ~\ref{LIR-IR}c for the sample where the IR sources 
have been removed).
The first point is that the difference between the ``LIRG and ULIRG'' and ``all galaxy'' or ``all galaxy - IR'' local environment distributions is no longer significant since the KS tests give the probability that the ``LIRG and ULIRG'' local environment distribution comes from the ``all galaxy'' or ``all galaxy without IR''one is within 1$\sigma$ . However, this magnitude cut might not be entirely satisfactory since some residual magnitude trends might remain in the subsamples and could bias the result. We have then used the ``all galaxy'' sample to generate $\sim$ 100 subsamples with the same number of sources as 
the ``LIRG and ULIRG '' one and that reproduce the magnitude distribution of LIRGs and ULIRGs. We obtain a KS probability of 0.73$_{-0.26}^{+0.21}$ that the ``LIRG and ULIRG'' local environment distribution and the Monte-Carlo generated ones come from a similar parent population, so there are no longer significant differences. We also use the ``all galaxy sample without IR sources'' in a similar way and obtain KS=0.60$_{-0.18}^{+0.25}$.
As a consequence, the trend that LIRGs and ULIRGs are in denser environments is substantially weakened when considering sources with similar luminosities. Thus, the LIRGs and ULIRGs tend to have similar local environments as sources 
located in the same luminosity range. 
This shows that LIRGs and ULIRGs do not tend to be peculiar objects in the local over-density - M$_B$ relation.

The second result is that the local environment of LIRGs and ULIRGs still remains significantly denser than that of the ``field'' 
and less dense than the ``group'' samples. This trend remains valid with a magnitude selection as well as for ``all galaxy'' and ``all galaxy without IR'' samples .

\subsection{LIRGs and ULIRGs and the color - magnitude relation.}

\begin{figure}
       \resizebox{\hsize}{!}{\includegraphics{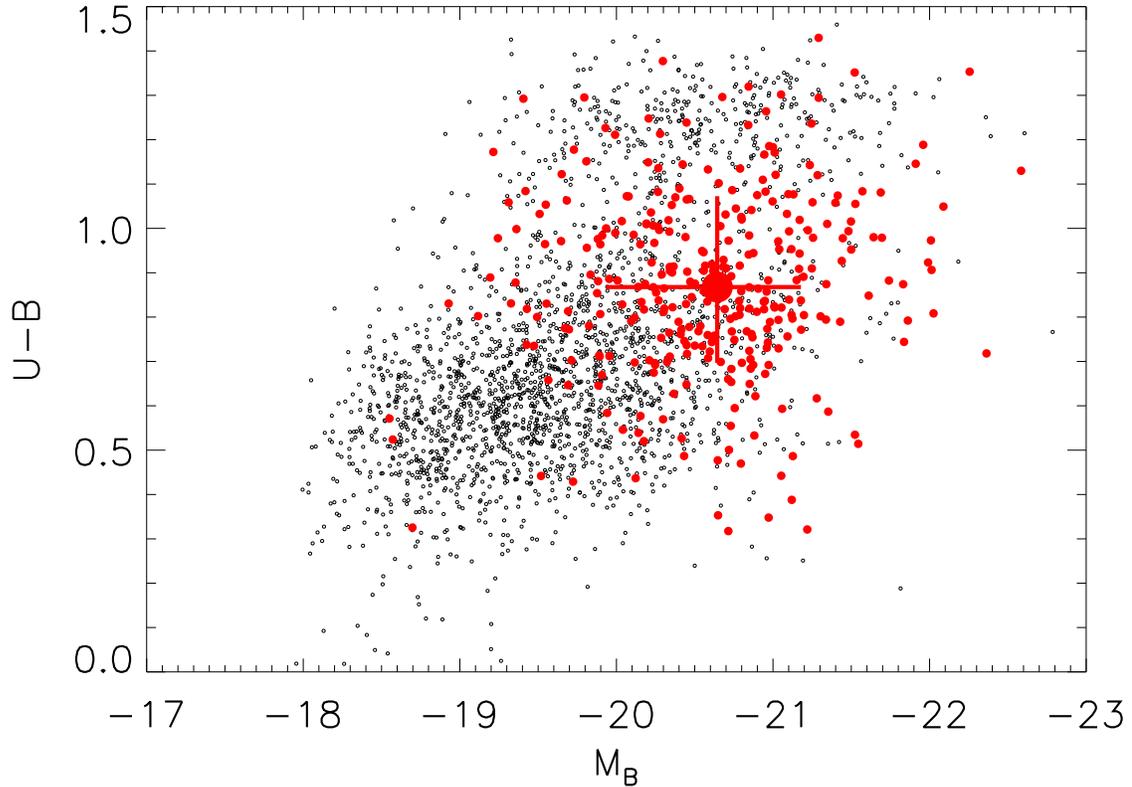}}    
\caption {(U-B) color as a function of M$_B$. The black open circles represent galaxies in a similar redshift range as the LIRGs and ULIRGs while the red plain ones are the LIRGs and ULIRGs. The big red dot is associated with the median magnitude and color value for LIRGs and ULIRGs with a 1$\sigma$ error bar. For the given mean ``LIRG and ULIRG'' luminosity, LIRGs and ULIRGs tend to have bluer colors than sources located in a similar redshift range. This can also be interpreted that for a given mean ``LIRG and ULIRG'' color, LIRGs and ULIRGs tend to be optically brighter.  [{\it{See the electronic edition of the Journal for a color version of this figure.}}]}
\label{cmdd}   
\end{figure}

Although we had previously found that LIRGs and ULIRGs lie in denser mean environments 
than all the galaxies in a similar redshift range, in the two previous subsections, we saw that 
this trend is weaker when comparing to galaxies with similar color or magnitude. However,
within color- or magnitude-selected samples, the mean magnitude - color behavior of 
LIRGs and ULIRGs differs from that of the other galaxies. 
This can be seen in Figure~\ref{cmdd}, which presents the (U-B) color as a function of M$_B$; the 
black open circles represent all the galaxies in a similar redshift range as the LIRGs and ULIRGs 
while the red plain ones are the LIRGs and ULIRGs. The big red dot shows the median 
magnitude and color value for LIRGs and ULIRGs with a 1$\sigma$ error bar. For the given 
mean ``LIRG and ULIRG'' luminosity, LIRGs and ULIRGs tend to have bluer colors. This can also 
be interpreted that for a given mean ``LIRG and ULIRG'' color, LIRGs and ULIRGs tend to be bright in the optical. 

We have therefore tested our results against a color - magnitude selection to compare very similar 
optical sources. We used the ``all galaxy sample'' to generate $\sim$ 100 Monte Carlo subsamples 
with the same number of sources as the ``LIRG'' sample and the same color and  magnitude distribution. 
We obtained KS=0.90$_{-0.23}^{+0.06}$ that the two distributions come from the same population. 
We obtained KS=0.55$_{-0.22}^{+0.27}$ for the ``all galaxy without IR sample''. LIRGs and ULIRGs tend 
to have a very similar mean environment as sources with similar color and magnitude disregarding 
their IR activity. These results are very close to the ones obtained for luminosity selected 
samples as discussed in the previous subsection.

We also find that the local environment of LIRGs and ULIRGs still remains significantly denser 
than that of the ``field'' 
and less dense than the ``group'' samples. This trend remains valid with a color-magnitude 
selection as well as for ``all galaxy'' and ``all galaxy without IR '' samples.

\subsection{Studying LIRGs and ULIRGs in mass selected samples.}
It is difficult to interpret these results because they rely on color and luminosity, which might not be the best descriptors for LIRGs and ULIRGs when compared to non-IR sources. LIRGs and ULIRGs are intrinsically brighter 
mainly because they are forming young stars. Both the R-band selection of DEEP2 (rest-frame 3400 \AA $\,$) and the comparison at fixed M$_B$ will combine smaller blue galaxies with larger red galaxies. Therefore, we now test the dependence of our results against a mass selection. 

For the portion of the EGS imaged with J and Ks at Palomar, Bundy et al. (2006) estimated stellar masses using BRIJK photometry and synthetic SEDs from Bruzual \& Charlot (2003), which span a large range of star formation history and metallicities. To extend the stellar mass estimates to the whole EGS, we use the mean relation of Lin et al. (2007) derived from the Bell \& de Jong (2001) prescription. We find no systematic differences between masses estimated from Bundy et al. (2006) and Lin et al. (2007) for LIRGs and ULIRGs.  

\begin{figure}
       \resizebox{13cm}{!}{\includegraphics{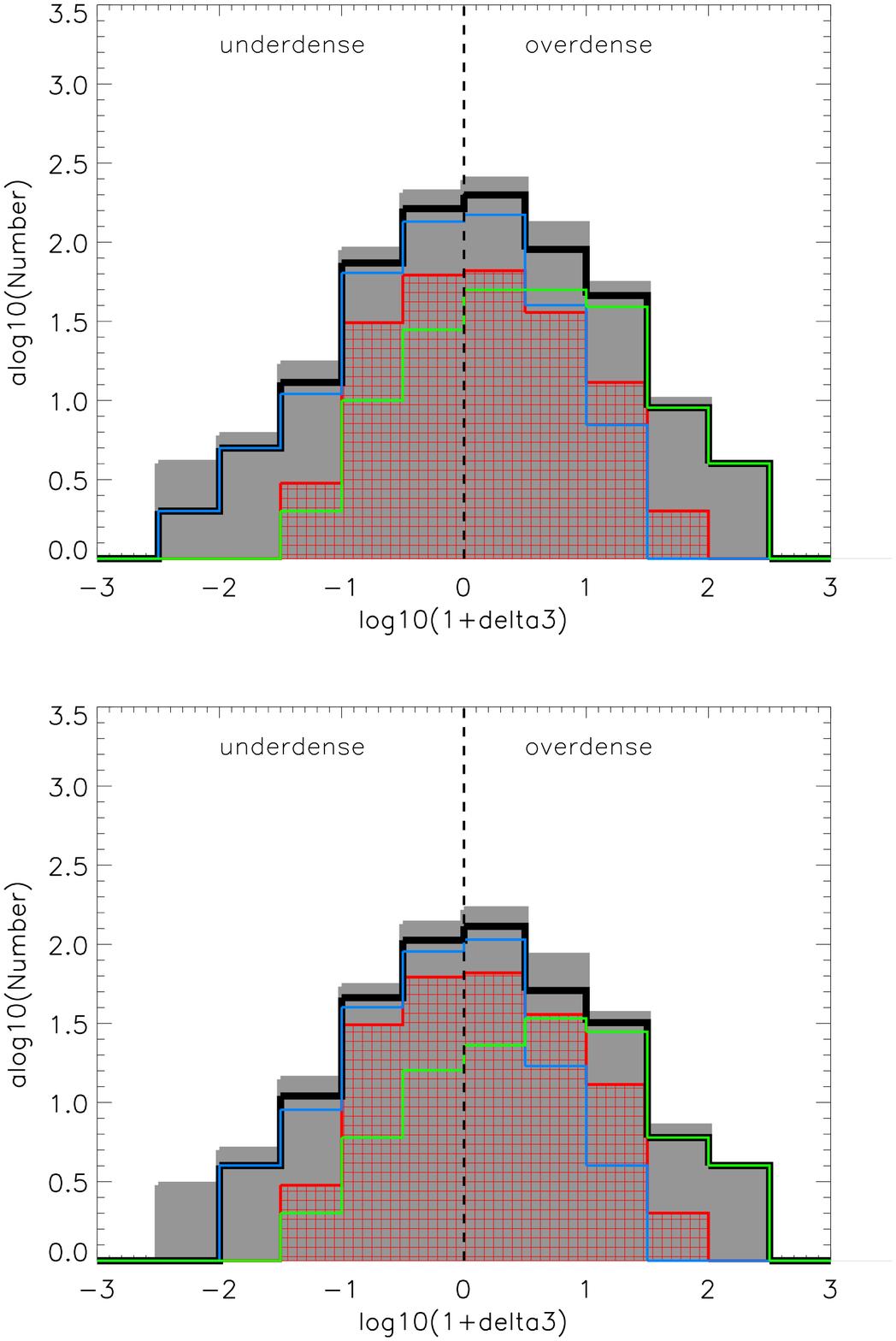}}
      \caption {Distribution of the local galaxy environment for the mass selected sample. {\bf{Top :}} Sample 
with IR galaxies (like in Figure 6). {\bf{Bottom :}} Sample with all sources not detected at 24 $\mu$m (like in Fig 7). }
\label{Masses}
\end{figure}

We restricted all samples to the 2-7 $\times$ 10$^{10}$M$_{\odot}$ range, which includes 68\% of the LIRG 
and ULIRG mass range. The results are presented in Figure~\ref{Masses}a and b for the normal samples and 
for the samples where IR sources have been removed. We generated 100 Monte Carlo subsamples for 
the ``all galaxy sample'' and ``all galaxy without IR sample'' to match the mass distribution of LIRGs 
and ULIRGs. We obtain a KS probability of 0.80$_{-0.26}^{+0.10}$ and 0.63$_{-0.22}^{+0.22}$ respectively 
that the LIRG and ULIRG local environment distribution and the generated ones come from a similar parent 
population. This behavior is very similar to that we find for the magnitude or magnitude - color selection. 
This is the consequence of the masses having been estimated  as a function of  both M$_B$ and U-B; however, 
the masses depend on M$_B$ to first order while the color correction is a second order effect 
(Bell et al., 2005, Lin et al., 2007). 

The local environment of LIRGs and ULIRGs still remains significantly denser than that of the ``field'' 
and less dense than the ``group'' samples. This trend remains valid with a mass selection as well as for the ``all galaxy'' and ``all galaxy without IR'' samples .

\section{What fraction of LIRGs and ULIRGs are classified as group members ?}
\label{frac_gr}
In contrast to the local clusters (Fadda et al., 2000), LIRGs and ULIRGs have been found in dense environments 
such as dense clusters and can be found in excess in some of them when compared to field sources 
at 0.5$ < $z$ < $0.9, (Geach et al., 2006, Marcillac et al., 2007, Bai et al., 2007). However the 
number of LIRGs and ULIRGs in very dense environments (clusters or outskirts of clusters) is small 
and not representative of the mean environment of the whole population. 
We now know that IR galaxies tend to have a mean local environment that is significantly denser than  
non group member sources and significantly less dense than ``group'' members; this statement has been tested against rest-frame colors and luminosity and mass. However we would like to estimate the fraction of sources that are group members.

Gerke et al. (2005) show that field sources represent roughly 67\% of the DEEP2 survey. They also 
represent roughly 73\% of the galaxies in the EGS. The remaining sources belong to groups. 
We have used the ``group member'' indicator as a way to test in an independent way the environment of LIRGs and ULIRGs.

Gerke et al. (2005), did not 
classify all sources in the EGS spectroscopic redshift catalog because the EGS sampling 
is slightly higher than in the three other fields studied in DEEP2. Since the group finding algorithm 
is optimized for a specific sampling, they re-sampled the EGS field data. Therefore, in this section we 
have restricted the four samples defined in section~\ref{envvv} to the subsamples of sources that 
have been analysed using the group-finding algorithm. In the ``all galaxy'' sample, 27\% of the sources 
are group members. This value decreases to 25\% if the MIPS sources 
are removed from the sample. In comparison, 33\% of the LIRGs and ULIRGs belong to groups. 
We performed a Monte Carlo simulation to test if this difference is significant.
We used the ``all galaxy''  sample to generate $\sim$ 10000 subsamples with the same number of sources as 
the ``LIRG and ULIRG'' one  and studied the distribution of the fraction of group members. Only 1.1\% the ``all galaxy'' 
sample have a ratio as high as the MIPS sample, which suggests that this difference is significant. 
The difference is even higher if we make the Monte Carlo simulation using the sample of ``all 
field without IR ''. 
This confirms that the IR sources lie more often in groups than the non-IR sources and agrees 
with local over-density estimator findings.

However, as seen in section~\ref{env}, the results concerning the local over-density estimators are 
strongly affected if we compare the ``LIRG and ULIRG'' 
environment to that of sources selected in color, luminosity or mass bins. We have therefore studied the fraction 
of {\it group} members in color, luminosity and mass selected samples to test this issue.
The results are presented in 
Table~\ref{gr}. We performed a Monte Carlo simulation to estimate the 
validity of each result. Although the fraction of LIRGs and ULIRGs that are group members tends to be 
higher when compared to 
sources with similar colors and luminosities, the Monte Carlo simulation shows that the difference is 
less significant than previously. All differences vanish when studying the mass selected samples. This 
confirms in an independent way the results found in the previous section. LIRGs and ULIRGs are more 
often found in groups compared to the general population of galaxies located in the same redshift range. 
However, this difference is weakened or vanishes when they are compared to a sample of galaxies with 
similar color, luminosity or mass.

\begin{table*}
\centering
\begin{tabular}{cccc}
\hline
\hline
 &LIRGs and ULIRGs & All galaxy & All galaxy - IR  \\    
\hline
no selection        & 33\% & 27\% & 25.5\%\\
Monte Carlo results &       & 1\%   & 0.1\%\\
\hline
(U-B) selected & 29\% & 25\% & 23\% \\
Monte Carlo results &     & 13\% & 2\% \\
\hline
M$_B$ selected     & 32\%& 28\% &26\%\\
Monte Carlo results &     & 13\% &3\%\\ 
\hline
Mass selected$^1$     & 32\%& 32\% &30\%\\

\hline
\hline
\end{tabular}
\caption{Fraction of sources being classified as group members as a function of color, luminosity and mass for the ``LIRG and ULIRG'' sample and the complete sample of sources with 0.7$\leq$z$\leq$1.} 

\noindent{\footnotesize {\it Comments} : (1) No Monte Carlo results are given since no trends are seen.} 

\label{gr}
\end{table*}

\section{Influence of the environment on LIRG and ULIRG activity at z$\sim$1.}

\subsection{Contribution of LIRGs and ULIRGs to SFR(IR) at z$\sim$1 as a function of environment}

In this section we examine the contribution to the total SFR of IR galaxies in different environments. 
We saw in section 6 that LIRGs and ULIRGs tend to  avoid very under-dense environments, but otherwise 
are found in a large diversity of environments at z$\sim$1, ranging from clusters (Geach et al., 2007, 
Marcillac et al., 2007), to groups or less dense structures. 
LIRGs and ULIRGs have been shown to play a significant role in galaxy evolution at z$\sim$1: 
Le Floc'h et al. (2005) show that the space density of LIRGs and ULIRGs is more than 100 times their 
density in the local universe. At z$\sim$0.7, they have the same contribution to the SFR as the star 
forming galaxies, and they dominate the SFR at higher redshift. 
A negligible fraction of LIRGs and ULIRGs belong to clusters or bigger bound structures; one reason is 
that these structures have a low number density at z$\sim$1. Therefore we have neglected their contribution. 
In Section~\ref{frac_gr}, we show that $\sim$30\% of LIRGs and ULIRGs are associated with groups in the 
EGS and this number weakly depends on color, luminosity and mass selection. 
At z$\sim$1, LIRGs and ULIRGs are responsible for $\sim$70\% of the comoving energy density 
(Le Floc'h et al. 2005); The contribution of LIRGs and ULIRGs that are  ``group members'' can be 
estimated to be 0.7$\times$0.3 $\sim$ 20\%, while the contribution from ``non group members'' 
is $\sim$ 50\%. This implies that 20 \% of the sources responsible for the IR star formation rate 
density and comoving energy density belong to groups in DEEP2 and are hosted by a dark matter halo 
with a lower limit mass of 6$\times$10$^{12}$M$_{\odot}$h$^{-1}$.

\subsection {Physical environment of IR-galaxies.}
\label{env_conc}

Here, we further characterize the environment of IR galaxies. We saw that LIRGs and ULIRGs tend to 
be in denser environments when compared to all R$\leq$24.1 galaxies and have an environment that is 
intermediate between blue and red galaxies; but this difference vanishes when their environment is 
compared to that of sources with similar color, luminosity or mass. Thus, LIRGs and ULIRGs are not 
outliers of the color-environment relations at z$\sim$1, since they have a similar environment 
as the other ``green valley'' galaxies. They also follow the luminosity-environment relations at z$\sim$1. 
LIRGs and ULIRGs always tend to be in a denser environment than ``field'' galaxies and less 
dense than ``group'' galaxies regardless of color, magnitude or mass selection. 

To go further, Coil et al. (2007) take advantage of the whole DEEP2 Galaxy Redshift Survey data to study 
the clustering properties of the ``green valley'' galaxies and show that they exhibit an intermediate 
clustering amplitude to red and blue galaxies. On large scales, they tend to have a clustering amplitude 
similar to that of red galaxies where at small scales, the amplitude is closer to that of blue galaxies. 
``Green'' galaxies are also found to have intermediate-strength ``fingers-of-god'' and have comparable 
kinematics to blue galaxies. They interpret these results as showing the ``green valley'' galaxies are in 
the same dense regions as the red galaxies but are less likely than red galaxies to lie at the cores of 
the overdensities. They suggest that ``green valley'' galaxies could reside on the outskirts of the same 
halos that red galaxies occupy. This interpretation is in good agreement with our results. It also 
agrees with the result of 
Cohen et al. (1999) and Gilli et al. (2007). Cohen et al. found that the MIR sources appear to be more 
clustered than the optical sources with $\sim$90\% of them lying within the statistically complete sample of 
redshift peaks. Gilli et al. found that LIRGs and ULIRGs in GOODS tend to be more clustered 
(r$_0$=5.1$\pm$0.8 h$^{-1}$Mpc) than the optically selected sources. Furthermore, this behavior is in 
agreement with the overall correlation between environment and SFR found at z$\sim$1 
(Cooper et al., 2007, Elbaz et al., 2007). 

It must be noted that very dense structure such as clusters or big groups have not been 
identified in EGS or GOODS-N and GOODS-S.  
Interestingly, distant (z$\sim$0.8) clusters show that LIRGs and ULIRGs are mainly found in the 
cluster outskirts and tend to have similar properties as the LIRGs and ULIRGs identified in the 
DEEP field surveys (Marcillac et al., 2007, Bai et al., 2007). 
This result appears to conflict with Elbaz et al. (2007), who found a structure at z=1.016 where the most 
vigorously star-forming galaxies are at the center of the groups. However, the results 
can be reconciled if the structure identified at z=1.016 is at an earlier stage of forming a cluster.

The tendency of LIRGs and ULIRGs to belong to the outskirts of structures such as groups may
suggest mechanisms responsible for the IR activity. Due to the incompleteness of the 
EGS spectroscopic redshifts, we can not exclude major mergers and strong galaxy--galaxy interactions from being an important 
mechanism to trigger IR activity. However, morphological studies have shown that mergers are 
not the only mechanism triggering LIRGs and ULIRGs at these redshifts (e.g., Lotz et al. 2006). 
If LIRGs and ULIRGs mainly belong to the outskirts of groups, galaxy-galaxy interaction such as 
minor mergers and pair interactions can play a role. The photometric and spectroscopic data used 
here are limited to R=24.1, which places a strong constraint on the upper mass limit of a companion that we
might detect that could be responsible for triggering IR-activity.
To estimate it, we have selected a sample of sources at R$\sim$24 at 0.7$ < $z$ < $1; they tend to have 
log(M)$\sim$ 10.0$_{-0.1}^{+0.3}$ for red galaxies and log(M)$\sim$ 9.3  $_{-0.3}^{+0.3}$ for blue ones. 
This mass range 
prevents us from studying minor mergers around IR galaxies, especially if the satellite 
is on the red side of the bimodality.
Thus, minor mergers and interactions with small satellites might be partly responsible for IR-activity. 
This is in agreement with morphology studies at z $\sim$ 1, which suggest that among the LIRGs 
classified as ``isolated spiral galaxies'', some show signs of asymmetry and disturbance that 
are too weak to be associated 
with a major merger (Shi et al. 2006). 

Weak interactions such as widely separated 
pairs can be another mechanism triggering IR activity. Lin et al.(2007) study the mid-IR properties 
and star formation of galaxy pairs and find a mid-IR enhancement for ULIRGs in pairs separated by 
less than a few tens of kpc. In addition, a smoother/weaker interaction between two galaxies with a 
larger separation could account for some enhancement of the star-formation activity. 
If the separation is large enough, the galaxies will not be classified as ``close pair'' sources 
but as two isolated galaxies that do not belong to bound structures. Thus, we cannot rule out from this 
study that the environment might play a role in IR activity through interactions 
such as minor mergers or by widely separated pairs. 

Do such interactions play a stronger role for IR galaxies than for sources of similar masses or luminosities? 
Or does the high IR luminosity have some ``duty cycle'' and all objects in overdense environments 
go through a LIRG or ULIRG phase (in which case, one can wonder if the environment is responsible for 
the IR activity, or just correlated with it) ?
It is also possible that the environment plays a very minor role in IR-activity, and that internal causes are responsible.


\section{Conclusion}

Using a combination of MIPS 24$\mu$m data with the multi-wavelength data available in the extended Groth strip (EGS), we investigate the role that the environment plays in the LIRG and ULIRG phenomena. Two environment indicators were used : the first one is a local environment estimator, the second is based on an algorithm that identifies groups and bound structures. These estimators are independent. We concentrate on IR-galaxies located at 0.7 $<$ {\it{z}} $<$ 1, where they dominate the IR luminosity function and star formation rate. 

Our results are as follows :

\begin{itemize} 
\item There is no trend between L$_{IR}$ and the local environment of LIRGs, showing that the environment is not gradually quenching or triggering the IR -- activity at z$\sim$1. 

ULIRGs tend to be in over-dense environments. However, they are also optically brighter than LIRGs and tend to have an environment comparable to sources with similar optical  luminosity.

\item  We find that LIRGs and ULIRGs have a mean local environment that is intermediate between those of blue and red galaxies.

They also tend to avoid very underdense (D $<$ -1) environments.

LIRGs and ULIRGs are found in 1.2 times denser mean local environments than 
that of other DEEP2 galaxies at similar redshift.

\item We find that the LIRG and ULIRG local environment differences decrease when compared to sources with similar rest frame (U-B) color and vanish when compared to sources with similar mass, M$_B$ or M$_B$-color distribution. Thus, LIRGs have similar environment properties as ``green valley'' galaxies; LIRGs and ULIRGs are not outliers of the color-environment and magnitude-environment relations at z$\sim$1. 

\item The local environment of LIRGs and ULIRGs still remains significantly denser than that of the ``field'' 
and less dense than the ``group'' samples. This trend remains valid with a magnitude, color, color magnitude or mass selection selection. 
These results are comparable to the picture given in Coil et al. (2007) where ``green valley'' galaxies could reside on the outskirts of the same halos that red galaxies occupy.

\item We find that about 30\% of LIRGs belong to groups with a minimum dark matter halo mass of 6$\times$10$^{12}$M$_{\odot}$h$^{-1}$;  They constitute 20 \% of the sources responsible for the IR star formation rate density and comoving energy density at z$\sim$1. 


\end{itemize}

\acknowledgements
This work is based on observations made with the {\em Spitzer} Space
Telescope, which is operated by the Jet Propulsion Laboratory,
California Institute of Technology under a contract with NASA
(contract number \#1407).  Support for this work was provided by NASA
through an award issued by JPL/Caltech (contract number \#1255094).

This paper makes use of DEEP2 and EGS data.
Funding for the DEEP2 survey has been provided by NSF grant AST05-07428 and AST05-07483.
Some of the data presented here were obtained at the W.M. Keck Observatory, which is operated as a scientific partnership among the California Institute of Technology, the University of California and the National Aeronautics and Space Administration. The Observatory was made possible by the generous financial support of the W.M. Keck Foundation. The DEEP2 team and Keck Observatory acknowledge the very significant cultural role and reverence that the summit of Mauna Kea has always had within the indigenous Hawaiian community and appreciate the opportunity to conduct observations from this mountain.

This paper makes use of photometric redshifts produced jointly by Terapix and the VVDS teams, and the CENCOS interface (http://cencosw.oamp.fr) was used for data retrieval and analyses.
This work is based on observations made with the {\em Spitzer} Space
Telescope, which is operated by the Jet Propulsion Laboratory,
California Institute of Technology under a contract with NASA
(contract number \#1407). Support for this work was provided by NASA
through an award issued by JPL/Caltech (contract number \#1255094).  

The authors would like to thank the anonymous referee for helpful comments. ESL acknowledges the support of STFC. M.C.C.\ acknowledges support provided by the {\it Spitzer
Space Telescope} Fellowship Program,
through a NASA grant administered by the Jet Propulsion
Laboratory at the California Institute of Technology. 
\nocite{*}
\bibliography{bib}

\end{document}